\documentclass{aa}
\usepackage[varg]{txfonts}

\usepackage{graphicx}
\usepackage{float}
\usepackage{lscape}
\usepackage{wrapfig}
\usepackage{epstopdf}
\usepackage{subfigure}
\usepackage{epstopdf}
\DeclareGraphicsExtensions{.ps}
\usepackage{booktabs}
\usepackage{adjustbox} 

\usepackage[]{natbib}
\usepackage{amstext}

\begin{document}

\title{{\it XMM-Newton} spectroscopy of the accreting magnetar candidate 4U0114$+$65 }
\subtitle{}

\author{Sanjurjo-Ferrr\'{i}n, G.$^{1}$, Torrej\'on, J.M.$^{1}$, Postnov, K.$^{2}$, Oskinova, L.$^{3}$, Rodes-Roca, J.J.$^{1}$, Bernabeu, G.$^{1}$ }
\institute{$^{1}$ Instituto de F\'isca Aplicada a las Ciencias y las Tecnolog\'ias, Universidad de Alicante, 03690 Alicante, Spain\\
$^{2}$ Sternberg Astronomical Institute, Moscow M.V. Lomonosov State University, Universitetskij pr., 13, Moscow 119234, Russia\\
$^{3}$ Institute for Physics and Astronomy, Universit\"{a}t Potsdam, 14476 Potsdam, Germany}

\abstract{}{4U0114$+$65 is one of the slowest known X-ray pulsars. We present an analysis of a pointed observation by the {\it XMM-Newton} X-ray telescope in order to study the nature of the X-ray pulsations
and the accretion process, and to diagnose the  physical properties of the donor's stellar wind.}{We analysed the energy-resolved light curve and the time-resolved X-ray spectra provided by the {\it EPIC} cameras on board {\it XMM-Newton}. We also analysed the first high-resolution spectrum of this source provided by the {\it Reflection Grating Spectrometer}.}{An X-ray pulse of $9350\pm 160$ s was measured. Comparison with previous measurements  confirms the secular spin up of this source. We successfully fit the pulse-phase-resolved spectra with Comptonisation models. These models imply a very small ($r\sim 3$ km) and hot ($kT\sim 2-3$ keV) emitting region and therefore point to a hot spot over the  neutron star (NS) surface as the most reliable explanation for the X-ray pulse. The long NS spin period, the spin-up rate, and persistent X-ray emission can be explained within the theory of quasi-spherical settling accretion, which may indicate that the magnetic field is in the magnetar range. Thus, 4U 0114$+$65 could be a  wind-accreting magnetar. 
We also observed two episodes of low luminosity. The first was only observed in the low-energy light curve and can be explained as an absorption by a large  over-dense structure in the wind
of the  B1 supergiant donor. The second episode, which was deeper and affected all energies, may be due to temporal cessation of accretion onto one magnetic pole caused by non-spherical matter capture from the structured stellar wind. The light curve displays two types of dips that are clearly seen during the high-flux intervals. The short dips, with durations of tens of seconds, are produced through absorption by wind clumps. The long dips, in turn, seem to be associated with the rarefied interclump medium. 
From the analysis of the X-ray spectra, we found evidence of emission lines in the X-ray photoionised wind of the B1Ia donor. The Fe K$\alpha$ line was found to be highly variable and much weaker than in other X-ray binaries with supergiant donors. The degree of wind clumping, measured through the covering fraction, was found to be much lower than in supergiant donor stars with earlier spectral types. 
}{The {\it XMM-Newton} spectroscopy provided further support for the magnetar nature of the neutron star in 4U0114$+$65. The light curve presents dips that can be associated with clumps and the interclump medium in the stellar wind of the mass donor.}


\titlerunning{X-ray spectroscopy of 4U 0114$+$65}
\authorrunning{Sanjurjo-Ferr\'{i}n et al.}

\maketitle

\section{Introduction}

\begin{figure}[ht!]
\centering
\subfigure{\includegraphics[trim={20 60 20 70},width=\columnwidth]{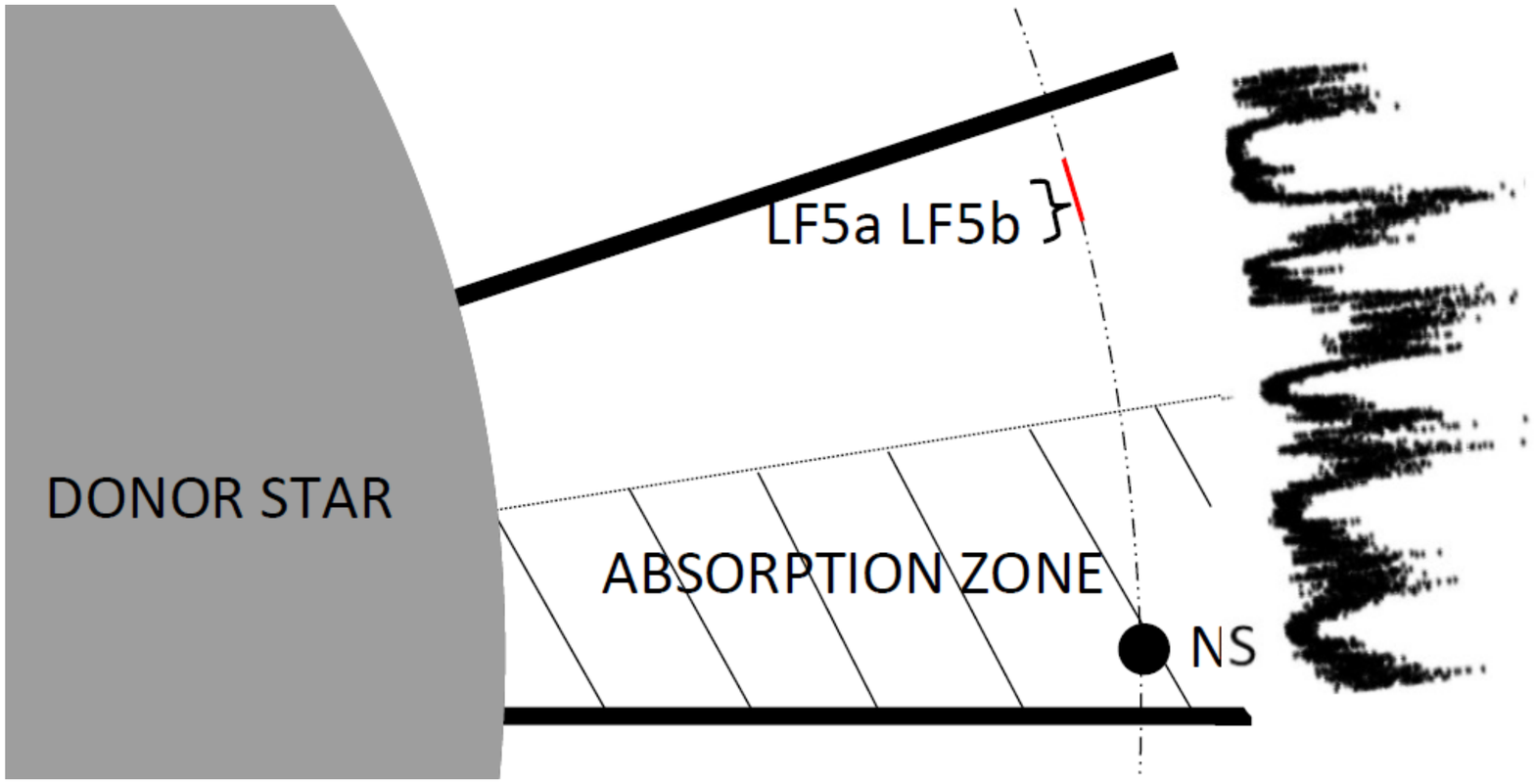}}
\caption{Sketch of the system where we can compare the size of the donor star with the orbit (NS size not to scale). The light curve is displayed parallel to the orbit. }
\label{orbita}
\end{figure}

High-mass X-ray binaries (HMXBs) are systems formed by a compact object (i.e. a neutron star or a black hole) orbiting a massive star (the companion). These are excellent laboratories where the process of accretion and the structure of the companion's stellar wind can be studied \citep{2017SSRv..tmp...13M}. The
source 4U 0114$+$65 was discovered by \citep{1977IAUC.3144....2D} in the SAS 3 Galactic survey. The donor star is V$^{*}$ V662 Cas, a luminous BI1a supergiant \citep{Reig} showing significant temporal and spectral variability over a wide range of timescales \citep{1985ApJ...299..839C}. With an orbital period of $\sim 11.6$ d, the neutron star (NS) orbits the donor deeply embedded into its wind at an orbital radius of $a \approx [1.34-1.65]R_{*}$ (see Table \ref{parameters} and Figure \ref{orbita}), thereby offering the opportunity to probe the inner wind of the B1 supergiant star. Other systems with similar donor spectral types but different orbital parameters are Vela X-1 (V$^{*}$ GP Vel, B0.5Ia) or 4U 1538$-$52 (QV Nor,  B0.5Ib).  

The source 4U 0114$+$65 is a non-eclipsing system (Pradham et al. 2015) and hosts an X-ray pulsar with a $\sim 2.6$ h spin period, which makes it one of the slowest known X-ray pulsars. In order to explain such a slow spin period, \citep{1999ApJ...513L..45L} suggested  that this object could have been born as a magnetar. Magnetars are neutron stars powered by their strong magnetic dipole fields, $B\simeq 10^{14}-10^{15} G$  \citep{1992ApJ...392L...9D}, and they can be present in HMXBs \citep{2008ApJ...683.1031B}. The pulse period of 4U 0114$+$65 has been changing fast over the years. It was first measured to be $\sim$ 2.78 h by \citep{1992A&A...262L..25F}. Eight years later, \citep{2000ApJ...536..450H} measured a period of $\sim$2.73 h, but in 2005, \citep{2005A&A...436L..31B} obtained a spin period of $\sim$2.67 h. A year later, \citep{2006MNRAS.367.1457F} obtained a spin period of $\sim$2.65 h. \citep{2011MNRAS.413.1083W} also discovered a spin period evolution from $\sim$ 2.67 h to 2.63 h between 2003 and 2008, with a spin-up rate of the neutron star of $\sim$ 1.09 $\times 10^{-6}s^{-1}$. In Table \ref{spinup} we compile all the spin periods measured to date.

Long periods of X-ray pulsars suggest high magnetic fields of accreting NS \citep{2013ARep...57..287I}. As an alternative, \citep{2006A&A...458..513K} proposed that the periodicities in this system could be explained by the accretion of a structured wind produced by tidally driven oscillations in the B-supergiant star photosphere induced by the closely orbiting NS.

\begin{table}[ht]
\centering
\begin{adjustbox}{max width=\columnwidth}
\begin{tabular}{rrr}
\toprule
Companion & & \\
\midrule
Spectral Type & B1Ia & \citet{Reig} \\
T$_{\rm eff}$ (K) & $24000\pm3000$ & \citet{Reig} \\
Radius ($R_{\odot}$) & $37\pm15$ & \citet{Reig} \\
Mass ($M_{\odot}$) & $16\pm5$ & \citet{Reig} \\
& & \\
\midrule
System & & \\
\midrule
$M_{V}$ & $-7\pm1$ & \citet{Reig} \\
$d$ (kpc) & $7\pm3$ & \citet{Reig} \\
$v \sin i$ & $96\pm20$ & \citet{Reig} \\
$E(B-V)$ & $1.24\pm0.02$ & \citet{Reig} \\
$BC$ & $-2.3\pm0.3$ & \citet{Reig} \\
$M_{\rm bol}$ & $-9.3\pm1$ & \citet{Reig} \\
$L_{X} (\times 10^{36} \rm{erg} \ s^{-1}) $ & $1.37_{-0.21}^{+0.24}$ & This work \\
$P_{\rm orb}$ (days) & $11.6\pm0.1$ & \cite{grun} \\
$P_{\rm superorb}$ (days) & $30.7\pm0.1$ & \citet{2006MNRAS.367.1457F} \\
$P_{\rm pulse}$ (s) & $9350\pm160$ & This work \\
$v_{\infty}$ (km/s) & 1200 & \citet{Reig} \\
$v_{wind}$ (km/s) & $500 \pm  80$ \\
$v_{orb}$ (km/s) & $244 \pm 24 $\\
$v_{rel}$ (km/s) & $560 \pm 70 $\\
$a$ ($R_{*}$) &$ 1.51 \pm 0.16 $\\
$R_{Bondi}(\times 10^{11}\mathrm{cm}) $ & $1.2 \pm 0.3   $   \\
\bottomrule
\end{tabular}
\end{adjustbox}
\caption {Velocity of the stellar wind, calculated with a $\beta$-law equation: $v_{wind}=v_\infty(1-\frac{R_*}{a})^\beta$, with the value $\beta=0.8$ \citep{1999isw..book.....L}, at the site of the NS. $R_{*}$  refers to the donor star radius.}
\label{parameters}
\end{table}

In this paper we present an analysis of a 49 ks {\it XMM-Newton} pointed observation, which represents the longest uninterrupted observation performed so far for this source using modern CCD\footnote{Charge-coupled device} X-ray telescopes.

\section{Observations and analysis}

We observed 4U 0114$+$65 for 49 ks using the X-ray Multi-Mirror Mission ({\it XMM-Newton}), launched by the European Space Agency (ESA) on December 10, 1999. {\it XMM-Newton} carries three high throughput X-ray telescopes and one optical monitor. The three {\it EPIC} instruments contain imaging detectors covering the energy range 0.15–10 keV with moderate spectral resolution. The two {\it EPIC MOS} cameras and the {\it EPIC} {\it PN} instrument were operated in small-window mode in order to avoid any pile-up. A thin filter was used to avoid optical load.

The observations were carried out in the time interval between 21-05-2015 20:43:19 UT $MJD= 57163.8634$ and 22-05-2015 10:01:53 UT, corresponding to an average orbital phase $\phi\approx 0.7$  \citep[using a $P_{\rm orb}=11.5983$ d and the ephemeris of][]{2015MNRAS.454.4467P}, which covers approximately $5\%$ of the orbit. Our {\it XMM-Newton} observation was performed well outside the minimum in the long-term light curve (see Pradham et al. 2015 for {\it Suzaku} observations performed during the minimun). The data can be found in the XMM archive under the ID \textit{0764650101}. The {\it EPIC} data were reduced using the \textit{Science Analysis Software v5.3.3 (SAS)}. The data were first processed through the pipeline chains and filtered. For {\it EPIC MOS}, only events with a pattern between 0 and 12  were considered, filtered through \#XMMEA EM. For {\it EPIC} {\it PN}, we kept events with flag=0 and a pattern between 0 and 4 \citep{2001A&A...365L..18S}. We also checked whether the observations were affected by pile-up using the task \texttt{epatplot,} with negative results. For the spectral analysis we combined the data of the three {\it EPIC} cameras ({\it MOS1}, {\it MOS2,} and {\it PN}) into one single spectrum using the task \texttt{epicspeccombine}. The spectra were analysed and modelled with the \textit{XSPEC} package.\footnote{maintained by HEASARC at NASA/GSFC.} The emission lines were identified thanks to the \textit{ATOMBD\footnote{http://www.atomdb.org/}} data base. On the other hand, the light-curve timing analysis was performed only on the {\it PN} data because of its higher time resolution. The photon arrival times were transformed to the solar system barycenter. To analyse the light curves, we used the \texttt{period} task inside the \textit{Starlink} suite\footnote{http://starlink.eao.hawaii.edu/starlink}.

The Reflection Grating Spectrometer {\it RGS} data were processed through the \texttt{rgsproc} pipeline SAS meta-task. This routine determines the location of the source spectrum on the detector, and performs the extraction of the source spectrum as well as of the background spectrum from a spatially offset background region. Finally, the task computes tailored response matrices for the first and second order. Both orders for {\it RGS} 1 and 2 were combined for the spectral analysis, providing a high-resolution spectrum in the energy range from 0.33 to 2.13 keV for the first time.

\section{Results}

\subsection{Light curve and pulse period}

\begin{figure*}[ht]
\centering
\subfigure{\includegraphics[trim={20 60 20 70},width=\textwidth]{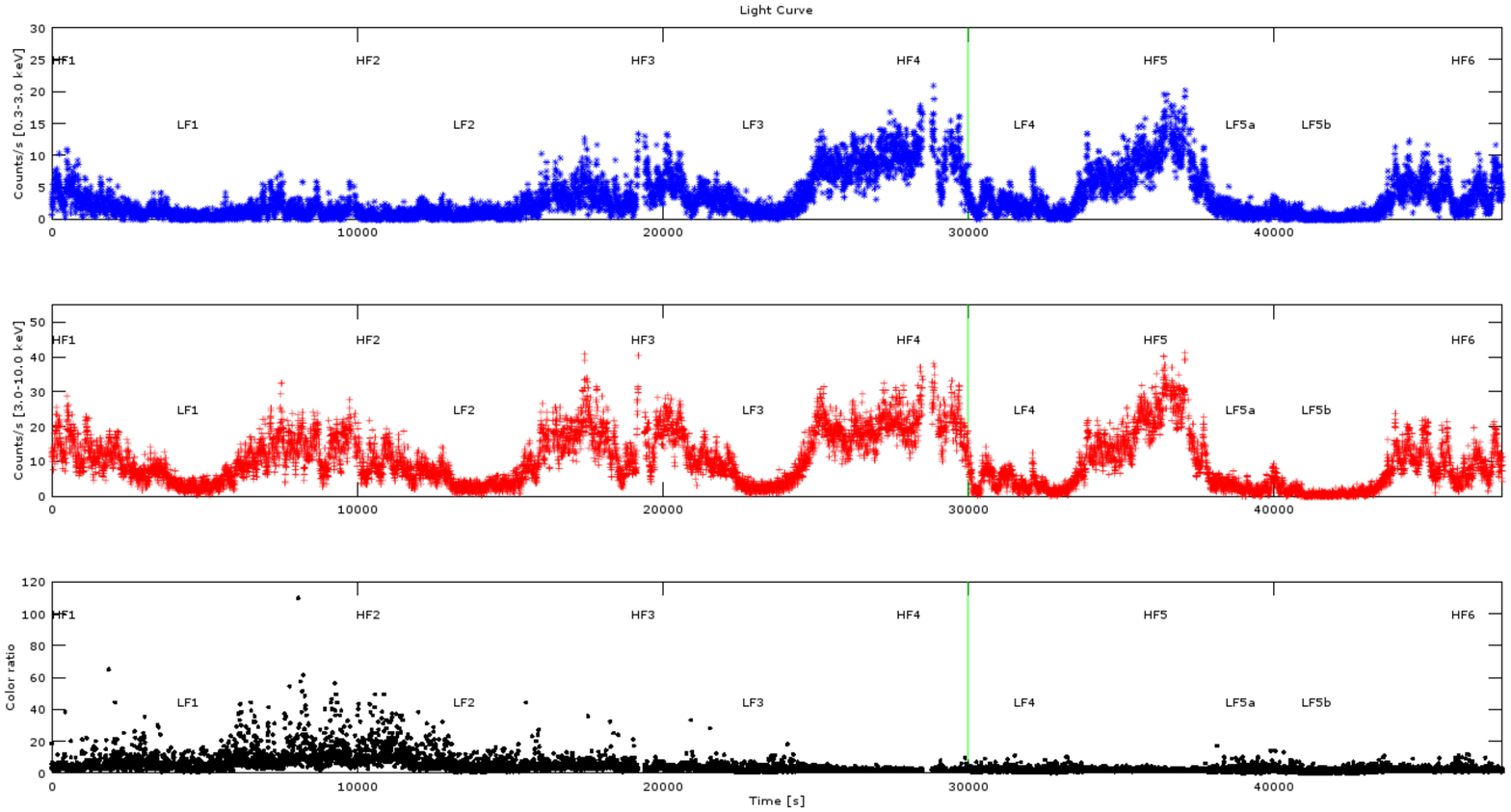}}
\caption{Light curve in two different energy ranges: upper panel 0.3-3.0 keV, and middle panel 3-10 keV. The lower panel shows the colour ratio (3-10)keV/(0.3-3.0)keV. The vertical green line divides T1 and T2 time intervals. The labels LF5a and LF5b refer to low-flux intervals, see text for detail. The epoch time corresponds to 57255.50 MJD}
\label{lctotal}
\end{figure*}

\begin{table*}[ht]
\centering
\caption{Compilation of spin period measured from 1992 until the present. To calculate the orbital  phase, we used as the initial epoch $T_{0}=55763.0;$ the value was taken from \citet{2015MNRAS.454.4467P}}
\begin{adjustbox}{max width=\textwidth}
\begin{tabular}{rrrrrrr}
\toprule
MJD Beginning& MJD End & Inital Phase & Final Phase& Period (s) & Author & Mission \\
\midrule
46431.0 & 46432.0 &0.36 & 0.44 &$10010\pm40$ &\citet{finley}& EXOSAT\\
50357.0 & 50387.0 &0.86 & 0.44 &$9830\pm40$ & \citet{2000ApJ...536..450H}&RXTE\\
53345.6 & 53346.8 &0.53 & 0.63 &$9605\pm14$ & \citet{bonning}&INTEGRAL\\
53347.8 & 53349.8 &0.72 & 0.89 &$9605\pm14$ & \citet{bonning}&INTEGRAL\\
53505.0 & 53535.0 &0.27 & 0.86 &$9540\pm40$ & \citet{2008MNRAS.389..608F}&PCA/HEXE\\
53717.0 & 53747.0 &0.55 & 0.14 &$9518\pm11$ & \citet{2008MNRAS.389..608F}&PCA/HEXE\\
52985.0 & 52987.0 &0.44 & 0.61 &$9612\pm20$ & \citet{2011MNRAS.413.1083W}&INTEGRAL/IBIS\\
53042.0 & 53044.0 &0.36 & 0.53 &$9600\pm20$ & \citet{2011MNRAS.413.1083W}&INTEGRAL/IBIS\\
53353.0 & 53355.0 &0.17 & 0.34 &$9570\pm20$ & \citet{2011MNRAS.413.1083W}&INTEGRAL/IBIS\\
53553.0 & 53556.0 &0.41 & 0.67 &$9555\pm15$ & \citet{2011MNRAS.413.1083W}&INTEGRAL/IBIS\\
53724.0 & 53726.0 &0.16 & 0.33 &$9520\pm20$ & \citet{2011MNRAS.413.1083W}&INTEGRAL/IBIS\\
54579.0 & 54581.0 &0.87 & 0.05 &$9475\pm24$ & \citet{2011MNRAS.413.1083W}&INTEGRAL/IBIS\\
55763.4 & 55764.6 &1.00 & 0.10 &$9370 \pm60$ & This work (2011) &{\it Suzaku}\\
57163.9 & 57164.4 &0.74 & 0.79 &$9350\pm 160$ & This work (2016)&{\it XMM-Newton}\\
\bottomrule
\end{tabular}
\end{adjustbox}
\label{spinup}
\end{table*}

The low-energy (0.3-3 keV) and high-energy (3.0-10 keV) light curves of the source are presented in Figure \ref{lctotal} along with their ratio. The light curves clearly show a main pulse,
in addition to which the stochastic variability, typical of wind accretion, can be seen, revealing the complexity of the accretion process. There is a strong pulse-to-pulse variability. The three and a half first pulses are similar in duration and in shape, whereas the last two pulses are very different. From now on, we refer to the three first pulses as T1 and to the last two pulses as T2. 

We performed time-resolved spectroscopy to divide the light curve into time bins corresponding to high flux (HF) and low flux (LF) states, respectively. These intervals are shown in Figure \ref{lctotal} in chronological order (note that LF5 is additionally divided into LF5a and LF5b because LF5 is much longer than the other low-flux states). The duration of the low and high states is not constant (see Figure \ref{lctotal}).

The light curve is complex, therefore  we used only the \textit{T1} interval and restricted the analysis to the high-energy range, from 8 keV to 10 keV, where the photoelectric absorption effects would be negligible, to compute a reliable spin
period. We calculated the spin period using several techniques.
The Lomb-Scargle method redefines the classical periodogram in such a manner as to make it invariant to a shift of the origin of time \citep{1976Ap&SS..39..447L,1982ApJ...263..835S}. The Fourier method\texttt{} performs a classical discrete Fourier transform on the data and sums the mean-square amplitudes of the result to form a power spectrum. The \texttt{Clean}  algorithm basically deconvolves the spectral window from the discrete Fourier power spectrum (or dirty spectrum), producing a clean spectrum, which is largely free of the many effects of spectral leakage \citep{1987AJ.....93..968R}. Finally, we calculated the time separation between consecutive low states as another measure of the spin period. We chose the low states because they show less variability than the high states. First we selected the time intervals with fewer than 6.5 counts/s, then we fitted these data to a Gaussian distribution, calculated the mean number of counts/s and rejected all of the counts beyond the two-sigma range. Then we rejected the outliers outside of the low states following the same method. The results are shown in Table \ref{perstar}. Clearly, all the results are consistent. The associated errors are relatively large because the light curve covers only 3.5 cycles of the pulse period. 

\begin{table}[htbp]
\centering
\caption{Pulse period obtained with different methods.}
\begin{tabular}{rr}
\toprule
Method& Period (s) \\
\midrule
\texttt{Lomb-Scargle} &$ 9400\pm 300$\\
\texttt{Fourier} & $9200\pm 700$ \\
\texttt{Clean} & $9400\pm 700$ \\
& \\
\texttt{Low-to-low} & $9350\pm200$ \\\\
\textit{Weighted average and $\sigma$ }& $9350\pm160$ \\
\bottomrule
\end{tabular}
\label{perstar}
\end{table}

\begin{figure}[ht]
\includegraphics[trim={0 7cm 0 7cm},width=\columnwidth]{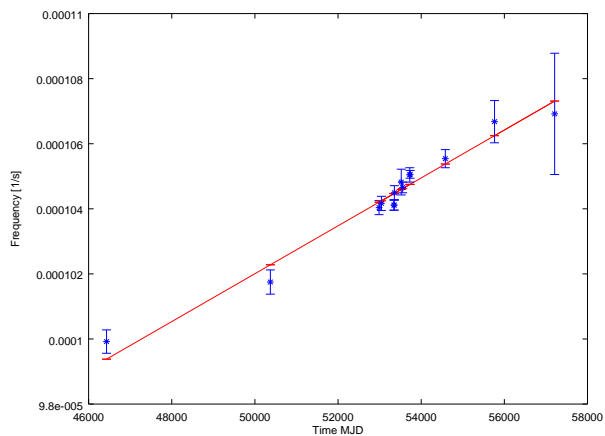}
\caption{Evolution of the pulse frequency from 1986 to 2015. }
\label{spinupf}
\end{figure}

\begin{figure}[]
\centering
\subfigure{\includegraphics[trim={2cm 7cm 2cm 7cm},width=\columnwidth]{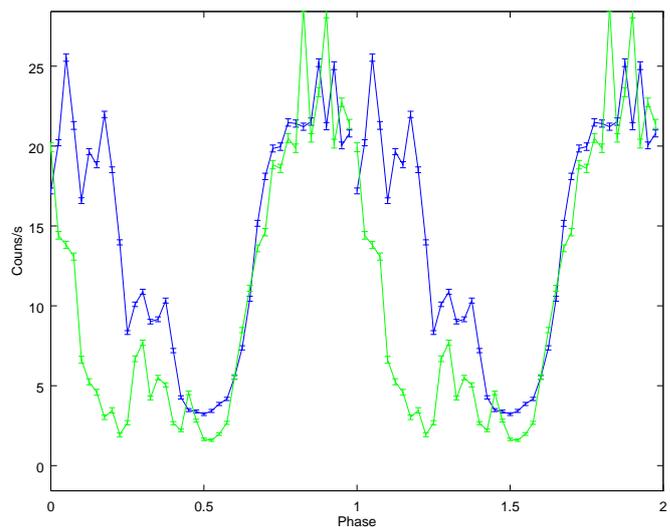}}
\caption{0.3-10 keV light curve,  folded on the spin period ($9320\pm1 s$), for T1 (blue solid line) and T2 (green dashed line). The epoch time is 57255.50 MJD.}
\label{foldedpulse}
\end{figure}

We finally computed the weighted average of pulse periods and determined a spin period of $9350\pm 160$ s where the uncertainty corresponds to the weighted error.  In order to check this result and increase the reliability of the pulse derivative, we also analysed a longer archival {\it Suzaku} light curve (2011-07-21 10:20:15, covering 12 cycles) using the same {\it Starlink} tools as described before. The derived period is given in Table \ref{spinup}. The historic period evolution is shown in Figure \ref{spinupf}.  The frequency derivative is $\dot{\nu}=(8.5  \pm 0.8) \times 10^{-15}$ Hz s$^{-1}$. Our observation is consistent with the secular spin-up trend of the pulsar \footnote{ Even though the source shows a clear spin-up trend, some episodes of torque reversal were detected, each lasting $\text{approximately one} $ year \citep[see][]{2006AdSpR..38.2779S}.}. We can also derive the pulse period at the time of our {\it XMM-Newton} observation using the historical trend. The result is $9320\pm1 s$, consistent with our previous determination (Table \ref{perstar}). Subsequently, we folded the light curve to the pulse period derived from the historical trend ( $9320\pm1 s$)  (Figure \ref{foldedpulse}). As expected, the pulse shape changes from T1 to T2. We compared the folded pulse with the
pulses obtained in previous studies that were performed at different orbital phases. Our T1 folded pulse (blue line in Figure \ref{foldedpulse}) is very similar to the folded pulse obtained by \citep{finley} (at $\phi_{\rm orb}=0.36$). On the other hand, our T2 pulse is similar to the pulses obtained by \citep{2006MNRAS.367.1457F}, whose observations covered all orbital phases (as they lasted 2.6 orbital cycles each). The source shows pulse shape changes
that are unrelated with the orbital phase.

Finally, during the whole observation, but specially during the HF states, we identify sharp drops in the flux on rather short timescales, so-called dips. We carefully checked for any periodicity associated with these dips to ensure that they are not related to a potential short-pulse period. The periodograms, however, do not show any significant peaks. We conclude that the dips are stochastic in nature. 

\begin{figure}[ht]
\subfigure{\includegraphics[trim={25 300 70 0},width=0.9\columnwidth]{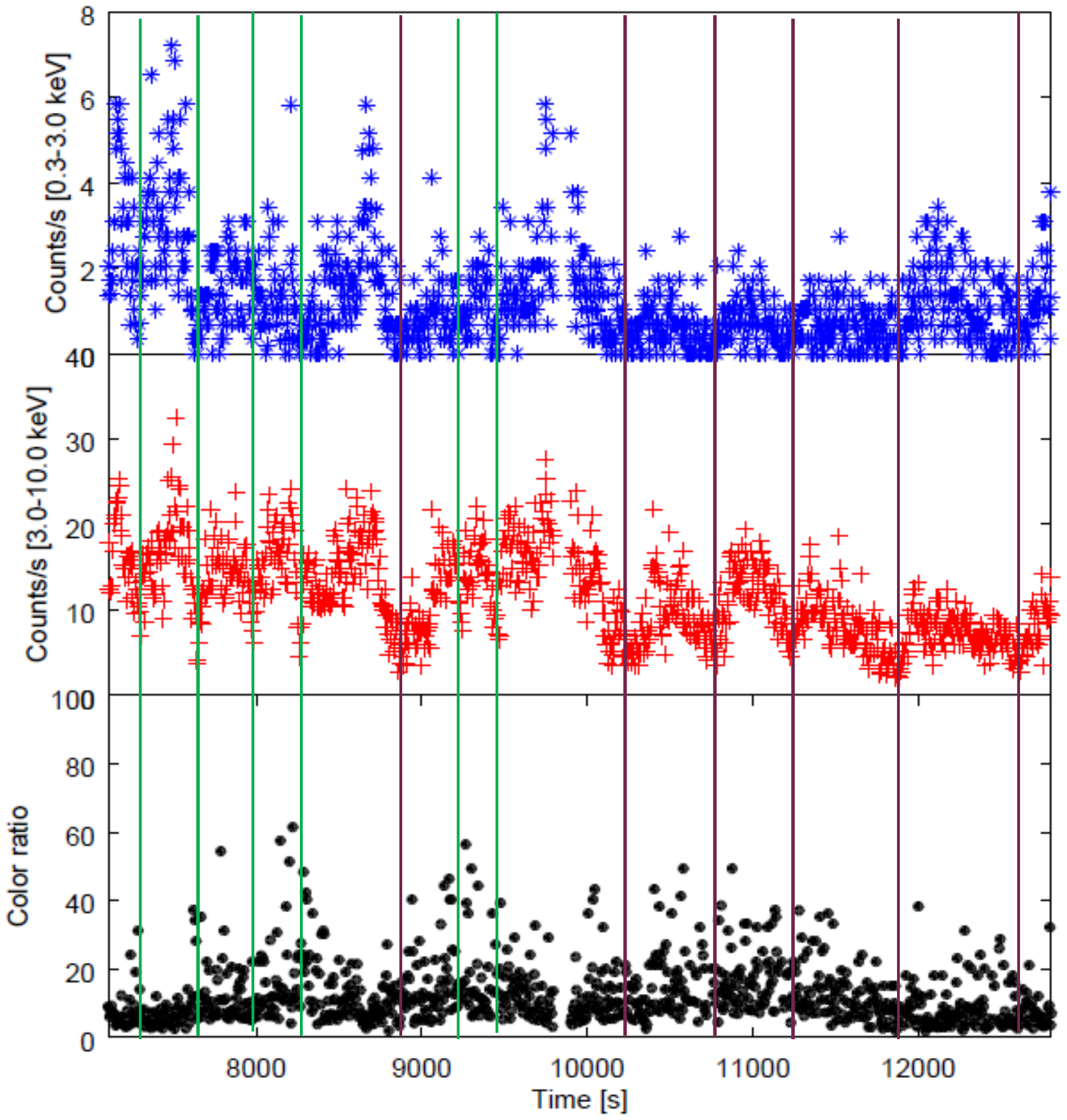}}
\subfigure{\includegraphics[trim={0 250 100 70},width=0.8\columnwidth]{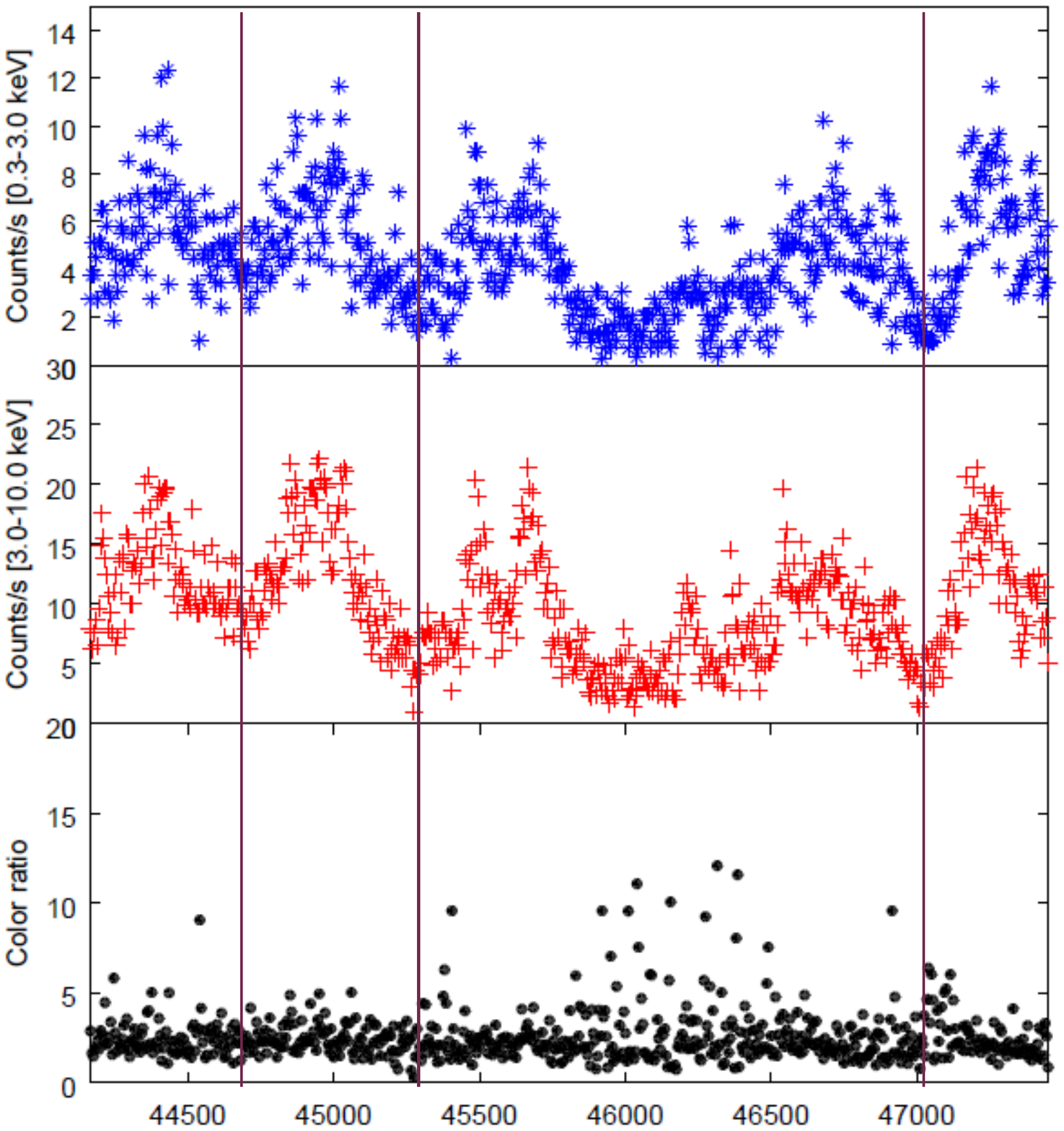}}
\caption{ Detailed view of two light curve sections, HF2 (top panel) and HF6 (bottom panel). Each panel is additionally subdivided into three divisions. The upper division corresponds to the low-energy light curve, the middle division to the high-energy light curve, and the lower division to the colour ratio. The short and long dips are marked by green and magenta lines, respectively.}
\label{HS2}
\end{figure}

 A close inspection shows that there are two types of dips: short, with typical durations of several tens of seconds, and long, lasting several hundreds of seconds. In Figure \ref{HS2} we mark them with vertical lines (green for short dips and magenta for long dips) for HF2 and HF6, where the dips are seen more clearly. The durations of the dips are $\sim 70 \pm 30$ s (short) and $\sim 530 \pm 110$ s (long). In the long dips the colour ratio tends to remain constant. Only two out of nine of the long dips show an increase in the CR. In the short dips, in turn, five out of six show a CR increase\footnote{Note that the colour ratio is much higher in HS2 than in HF6 because the higher general abosorption during the first part of the observation (see Table \ref{bmctable}).  

}

\subsection{{\it EPIC} spectrum}

We performed time-resolved spectroscopy for each HF and LF state seen in the light curve. The continuum was described using the  bulk motion Comptonisation (\texttt{bmc}) model. This\texttt{}  is an analytical model that describes the Comptonisation of soft seed photons by matter undergoing relativistic bulk motion \citep{1997ApJ...487..834T}. The model parameters are the characteristic black-body temperature of the soft photon source, a spectral (energy) index ($\alpha$), and an illumination parameter characterizing the fractional illumination of the bulk motion flow by the thermal photon source. 

The absortion is modelled by the Tuebingen-Boulder interstellar medium (ISM) absorption model  \texttt{Tbabs}. This model calculates the cross-section for X-ray absorption by the ISM as the sum of the cross-sections that are due to the gas-phase ISM, the grain-phase ISM, and the molecules in the ISM \citep{2000ApJ...542..914W}. 
The continuum is modified at low energies by a partial covering absorber in order to account for the wind clumping of the donor star, estimated through the partial covering fraction (parameter \textit{C}). Emission lines at 6.4 keV (Fe K$\alpha$ fluorescence), 0.78 keV (Fe \textsc{xviii}), 0.42 keV (N \textsc{vi} He-like), and 0.38 keV (C\textsc{v} Lyman $\alpha$) are clearly detected. We modelled them as Gaussians, with a fixed 
width of 50 eV, while their $norms$ were allowed to vary. The complete model is described by the equation

\begin{equation} \label{eq1}
\begin{split}
F(E) & = (C\exp(-N_{H}^{1}\sigma(E))+\\
& +(1-C)\exp(-N_{H}^{2}\sigma(E))) \left(BMC(E)+\sum_{i=1}^{4}(G_{i}) \right )
\end{split}
,\end{equation}

where $G_{i}$  represent the Gaussians describing the emission lines. In Table \ref{bmctable} we show the best-fit parameters. The reduced $\chi^2$ value is below 1.2 in most of the time bins, except in LF1, HS4, and HS5. In LF1 the high value of $\chi^2$  is due to instrumental features that are only present in the EMOS2 spectrum. On the other hand, the high $\chi^2$ value of HS4 and HS5 is due mostly to residuals around the Au edge. The high value of  $\chi^2$ of the average spectrum (Fig. \ref{smbb}) is expected because of the high variability of the source. 

\begin{figure}[h]
\centering
\subfigure{\includegraphics[width=\columnwidth]{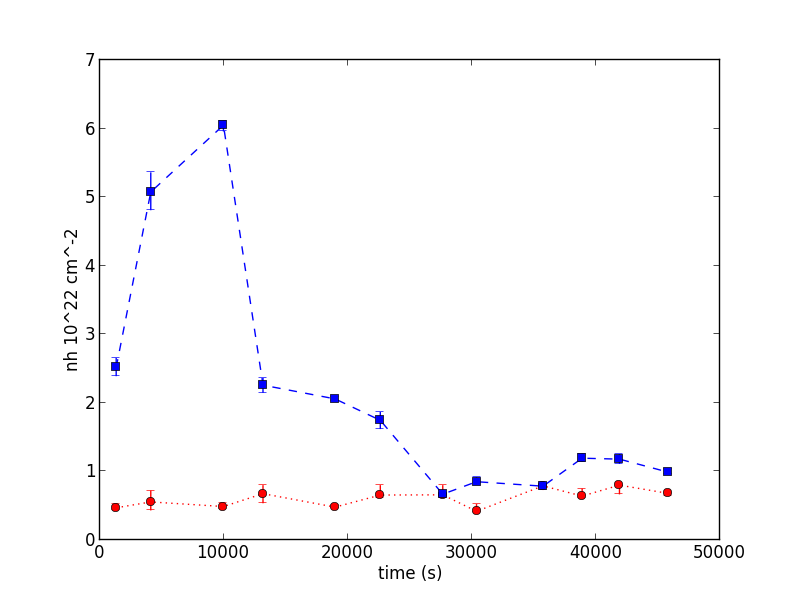}}
\subfigure{\includegraphics[width=\columnwidth]{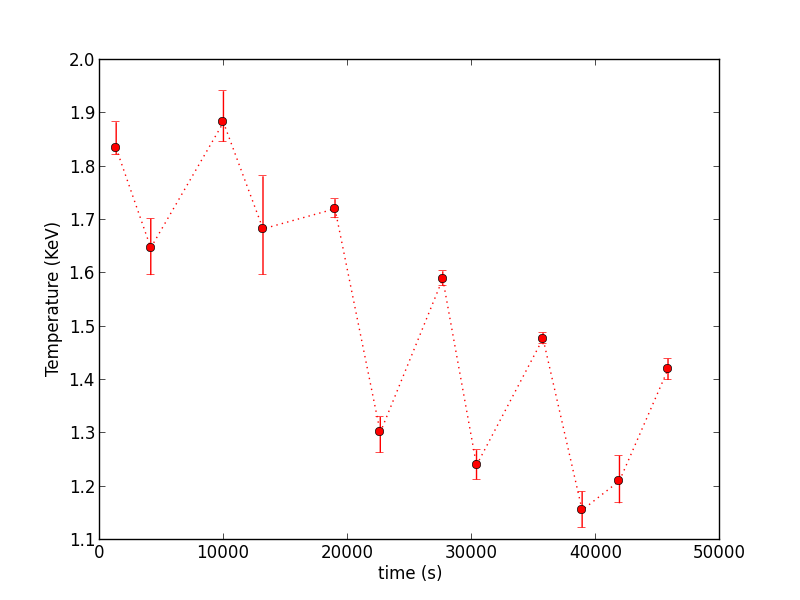}}
\caption {Evolution of the density columns (top) and radius of the seed soft photon source (bottom) for the \texttt{bmc} model. The top panel shows that the first-component absorption (red line) remains constant, compatible wih the ISM medium. The blue line represents the absorption of the ISM plus the circum-source environment, which changes during the observation. It is particularly high at the beginning of the observation, when the low-energy light curve is partially suppressed. }
\label{bmc}
\end{figure}

\begin{figure}[ht!]
\subfigure{\includegraphics[trim={80 20 0 300},width=\columnwidth]{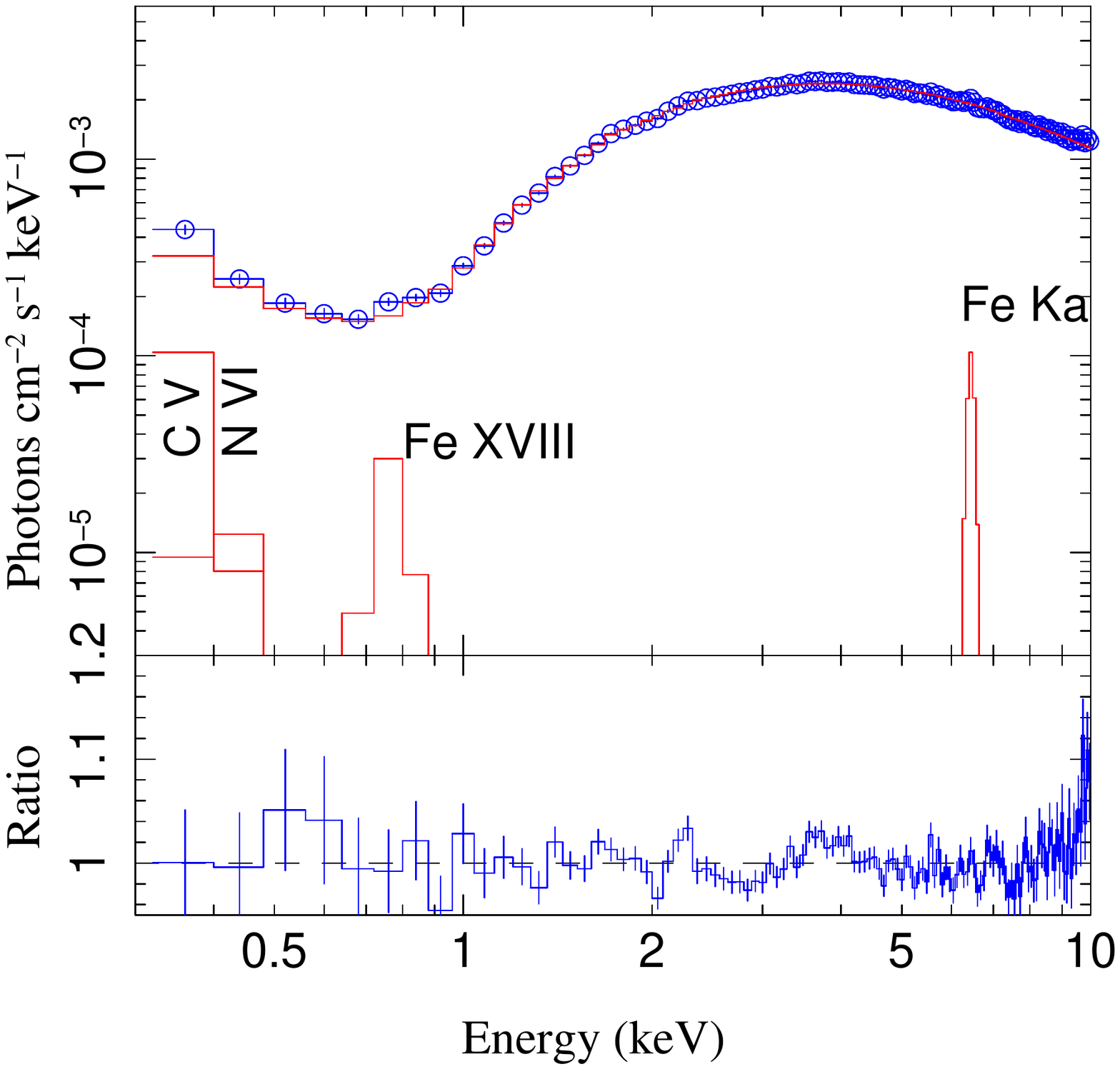}}
\caption{Average spectrum data (blue circles) with uncertainties (crosses) and fitted model (Eq. \ref{eq1}, red line). }
\label{smbb}
\end{figure}

In Figure  \ref{bmc} we present the variation of some parameters with time. The upper panel shows the evolution of the absorption.  Although \texttt{TBabs} is not devised to provide the absorption of X-ray radiation through ionised stellar winds, we use it here as a gauge to qualitatively estimate the changes in the absorbing column. The first absorption column stays constant at around $\sim 0.5\times 10^{22}$ cm$^{2}$ , which is compatible with the ISM value deduced from the optical data (0.8$\times 10^{22}$ cm$^{2}$, using the $E(B-V)$ of Table 1). In turn, the second absorption component represents the absorption of the ISM plus the circum-source environment, presumably the stellar wind. It is high at the beginning of the observation, coinciding with the suppression of the second peak in the light curve at low energies. Then it steadily decreases until the end of the observation. The partial covering fraction \texttt{C}, a proxy for the wind-clumping factor, is very low in general, compared to other sources with supergiant donors, which are always close to 1.   It is lower in the beginning of the observation, when the absorption is higher. These values are consistent with those found by \citep{2015MNRAS.454.4467P} using {\it Suzaku}. The spectral index $\alpha$  remains constant and <1, indicating an efficient Comptonising process (see Table \ref{bmctable}). The radius of the source emitting the seed soft photons, which are subsequently Comptonised, can be estimated assuming that the source is radiating as a blackbody of area $\pi R_{W}^{2}$ \citep{2004A&A...423..301T},
\begin{equation}
\label{ec:solucion}
R_{W}=0.6\sqrt{L_{34}}(kT)^{-2} [\rm km]
,\end{equation}

where $L_{34}$ is the luminosity of the \texttt{bmc} component in units of 10$^{34}$ergs/s and $kT$ is the temperature of the seed soft photons. In Table \ref{lum} we list the corresponding radii. The values are low ($\sim 1.5-3.5$ km), which is only compatible with a hot spot over a NS surface. 

\begin{table}[H]
\centering
\caption{Luminosity and radii of the seed soft photon source.}
\begin{tabular}{rrr}
\toprule
& $L_{X} (\times$ $10^{36}$) & $R_{W}$ \\
& (ergs/s) & km \\
\midrule
AS & $1.37_{-0.21}^{+0.24}$ & $2.17_{-0.03}^{+0.05}$ \\
HF1 & $1.4_{-0.6}^{+0.8}$ & $2.1_{-0.2}^{+0.3}$ \\
LF1 & $0.50_{-0.01}^{+0.01}$ & $1.56_{-0.01}^{+0.01}$ \\
HF2 & $1.91_{-0.01}^{+0.01}$& $2.34_{-0.01}^{+0.02}$ \\
LF2 & $0.80_{-0.02}^{+0.01}$ & $1.90_{-0.04}^{+0.05}$ \\
HF3 & $2.97_{-0.01}^{+0.01}$ & $3.49_{-0.01}^{+0.01}$ \\
LF3 & $0.25_{-0.01}^{+0.01  }$ & $1.78_{-0.01}^{+0.01}$ \\
HF4 & $3.01_{-0.01}^{+0.04}$ & $4.12_{-0.01}^{+0.01}$ \\
LF4 & $0.43_{-0.01}^{+0.01}$ & $2.57_{-0.01}^{+0.01}$ \\
HF5 & $3.47_{-0.01}^{+0.01}$ & $5.12_{-0.01}^{+0.01}$\\
LF5a & $2.15_{-0.03}^{+0.03}$ & $6.58_{-0.14}^{+0.16}$ \\
LF5b & $0.13_{-0.01}^{+0.01}$& $1.45_{-0.01}^{+0.01}$ \\
HF6 & $0.75_{-0.01}^{+0.01}$ & $2.56_{-0.05}^{+0.06}$\\

\bottomrule
\end{tabular}
\label{lum}
\end{table}

\begin{figure}[h]
\centering
\subfigure{\includegraphics[width=80mm]{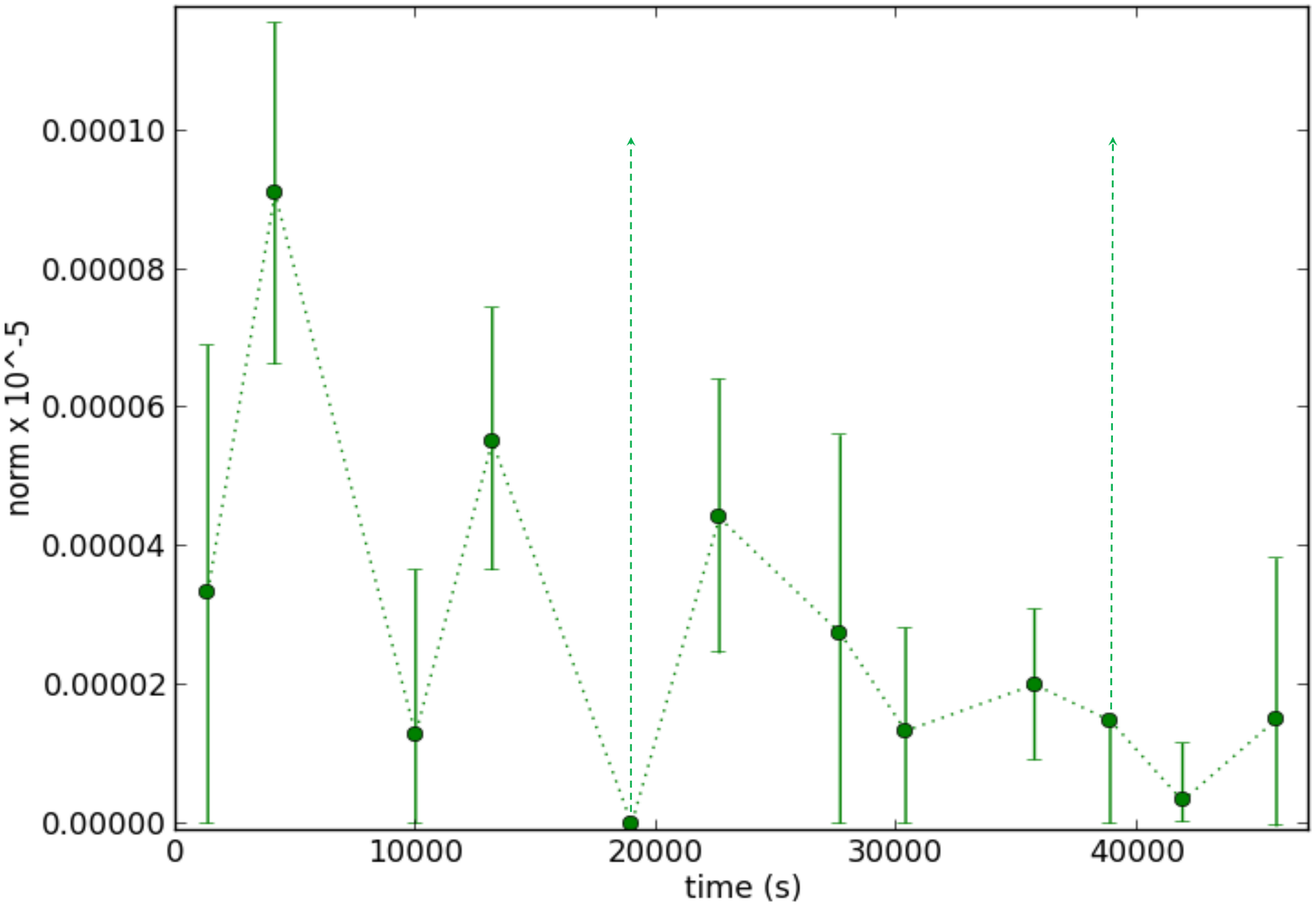}}
\subfigure{\includegraphics[width=80mm]{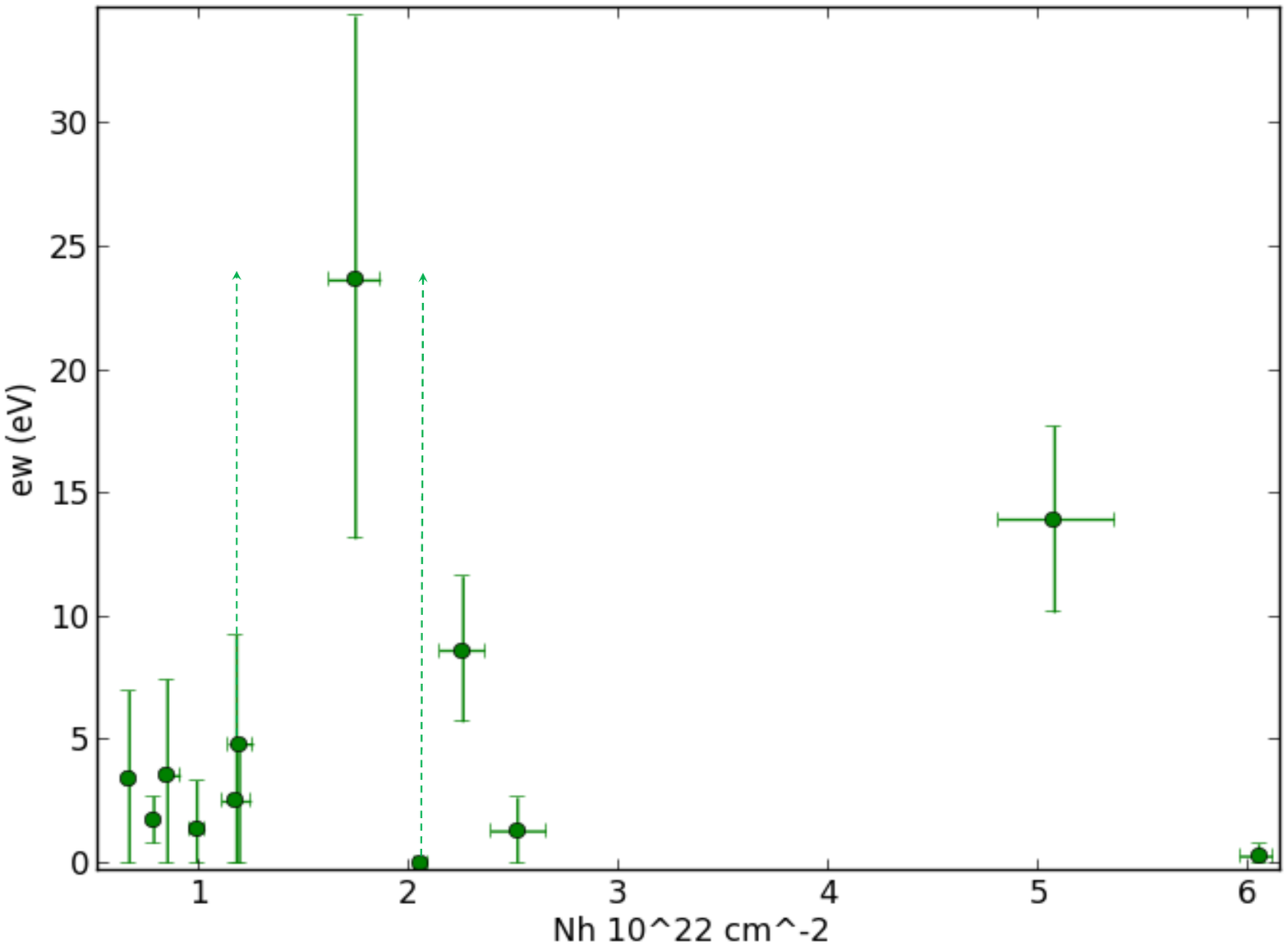}}
\caption {Variation of the Fe K$\alpha$ emission line. The upper panel shows the line flux versus time. The Fe line is weaker in the second half of the observation. In the lower panel, we plot the variation of the equivalent width ($EW$) versus the density column of the absorbing material. It does not show the linear trend that is expected when the reprocessing material is spherically distributed around the X-ray source \citep{2010ApJ...715..947T,2015A&A...576A.108G}. 
}
\label{emissionlines}
\end{figure}

\subsection{{\it RGS} spectrum}
\label{emlines}

\begin{figure}[H]
\subfigure{\includegraphics[trim={0 0 0 0cm},width=\columnwidth]{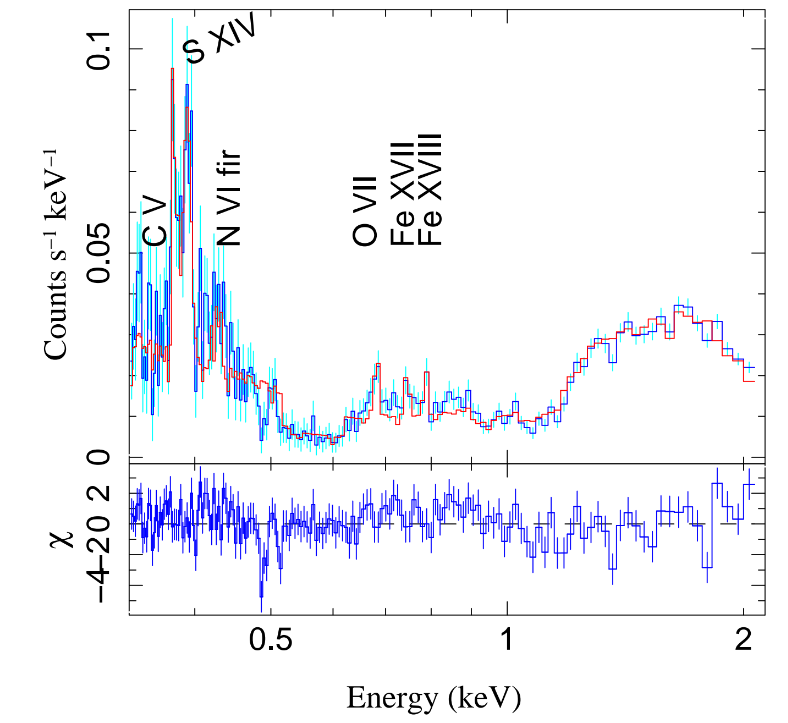}}
\caption{Time-averaged RGS spectrum data (blue) and model (red). }
\label{rgs}
\end{figure}

We present the first high-resolution spectra ever obtained for this source using the {\it XMM-Newton} {\it RGS} spectrometer in Fig. \ref{rgs}. The continuum was modelled following the EPIC analysis presented before. The spectral parameter $\alpha$ and the temperature $kT$ were fixed to the best-fit values obtained above\footnote{{\it RGS} is not sensitive to these parameters given the narrow energy range considered and its soft response.} , while the normalization and absorption were left free to vary. 

The emission lines were fitted with Gaussians. In agreement with the EPIC analysis, two broad emission features at 0.38 keV (C \textsc{v} Ly$\alpha$) and 0.42 keV (N \textsc{vi}) are clearly detected. When modelled using single Gaussians, they are very broad ($\sigma\sim 1$ keV), suggesting a complex structure. The N \textsc{vi} He-like triplet was modelled, thus, with three Gaussians representing the resonant ($r$), intercombination ($i$), and forbidden ($f$) transitions. The widths were fixed to $\sigma=50$ eV. The parameter $G=(f+i)/r= 4.1$ was found to be consistent with a photoionised plasma at $kT_{e}\sim 10^{6}$ K, while the parameter $R=f/i=0.56$ indicates electron plasma densities of $n_{e}\sim 10^{11}$ cm$^{-3}$ \citep{2000A&AS..143..495P}. The C \textsc{v} line is blended with a transition at 0.39 keV that
is compatible with S \textsc{xiv,} whose emissivity peaks at $3.2\times 10^{6}$ K. These lines dominate the spectrum at low energies. Aditional emission lines can be seen at 0.681 keV (O \textsc{vii} Ly$\alpha$), 0.746 keV (Fe \textsc{xvii}), and 0.791 keV (Fe \textsc{xviii}), although with a much lower intensity. 

\section{Discussion}
{\it The supergiant stellar wind}. Several emission lines are detected in the {\it EPIC} and {\it RGS} spectra. These lines are excited by the intense X-ray irradiation of the donor's wind. The strongest line is the Fe K$\alpha$ fluorescence line at 6.4 keV. However, the strength of this line ($EW_{\rm max}\approx 24$ eV) is much weaker than that seen in other HMXBs with supergiant companions \citep{2015A&A...576A.108G},\citep{2010ApJ...715..947T}. There are several possible explanations for this fact: a) the illuminated wind could have a much lower density, b) the wind material could be mostly ionised, or c) the wind could be under-abundant in Fe. Explanations b) and c) seem unlikely. A high ionisation of the wind material contradicts the fact that the photoelectric absorption of local origin is not negligible. An under-abundance of Fe is not expected at all in a system that underwent an SN explosion in the past. Explanation a) is also problematic. In a long-term study of the optical spectra, \cite{2016A&A...590A.122R} found a strong H$\alpha$ emission that sometimes shows a P Cygni profile. The donor star therefore has a substantial wind ($\dot M\sim 10^{-6}- 10^{-7} M_{\odot} {\rm yr}^{-1} $).  

The weakness of the Fe K$\alpha$ fluorescence line  might be related to the low covering fraction \texttt{C}. This would indicate a clumping factor much lower than those observed in other supergiant systems. In order to have a strong Fe K$\alpha$ line, the ionisation parameter $\xi=L_{\rm X}/n(r_{\rm X}) r_{\rm X}^{2}$, where $n(r_{\rm X})$ is the local particle density and $r_{\rm X}$ is the distance from the X-ray source (the NS), must be $\leq 10^{2}$.  In this case,  $\xi \geq 10^{2}$, which means that for an $L_{\rm X}\sim 10^{36}$ erg s$^{-1}$ and the characteristic distances within the system $r_{\rm X}\sim 10^{11-12}$ cm, the wind densities should be $n\sim 10^{10-12}$ cm$^{-3}$. 
 Wind densities like this are a factor 1 to 100 above the smooth wind density predictions. In other words, Fe K$\alpha$ is mostly produced in the wind clumps. Therefore, 4U 0114$+$65 seems to have the typical thick wind of a supergiant star, but with a much lower degree of clumping.  The donor in 4U 0114$+$65 (B1Ia) is the coolest of the supergiant X-ray binaries \citep{2015A&A...579A.111K}. With a $T_{\rm eff}=24$ kK, 4U 0114$+$65 is in the middle of the bistability jump \citep[see low panel in Figure 3 presented by][]{1999A&A...350..181V}. This jump is caused by the sudden change in the ionisation balance of Fe (\textsc{iii, iv}), which is a major contributor to the acceleration of the wind. We suggest that this could have a strong impact on the efficiency of the mechanism to form and/or destroy clumps. 

The light curve displays narrow (short) and wide (long) dips within the HF states (Figure \ref{HS2}). In the narrow dips the colour ratio tends to increase, which indicates increased absorption as the most probable cause. On the other hand, in the wide dips the colour ratio tends to remain constant. Thus, the origin of
the dips must be related to a general decrease in the mass accretion rate. A straightforward interpretation of the dips would indentify them with wind clumps (short dips) and the interclump medium (long dips). In order to estimate their characteristic sizes, we multiplied the time duration by $v_{rel}$ (Table \ref{parameters}).  The long dip size is $\simeq 1.6 \times 10^{-3} R_{*}$ ($4 \pm 3 \times 10^{9}$ cm), while the short dip size is $\simeq 1.2 \times 10^{-2} R_{*}$ ($2.9 \pm 2.2 \times 10^{10}$ cm) . Interestingly, the ratio long over short is $\simeq 7,$ which is close to the interclump to clump size ratio expected from theoretical models \citep{2017SSRv..tmp...13M}.

{\it The nature of the X-ray pulsations.} The data presented in this study support the interpretation of slow X-ray pulsations seen in the X-ray light curve of 4U 0114+65 in terms of the spin of accreting strongly magnetised NS. As we showed, the spectra of 4U 0114$+$65 can be described by Comptonisation models. The high temperature of the emitting plasma and the modest luminosity displayed by the source ($L_{\rm X}\sim 10^{36}$ erg s$^{-1}$) indicate a small emitting area (of the order of 3 km), compatible with a hot spot over the NS surface. Moreover, the secular spin period decrease observed along the years is not explained in the structured wind scenario presented by \citep{2006A&A...458..513K}. We therefore conclude that the most likely origin of the pulse is the spin of the NS. 

The observed properties of 4U 0114$+$65 can be explained in the frame of the theory of quasi-spherical settling accretion onto slowly rotating magnetised neutron stars elaborated by \cite{2012MNRAS.420..216S} and further developed in \cite{2014EPJWC..6402001S} (see \cite{2015ARep...59..645S} and \cite{2017arXiv170203393S} for a review of the theory and applications). This regime of accretion from a stellar wind can be realised in sources with moderate X-ray luminosity (below $\simeq 4\times 10^{36}$~erg~s$^{-1}$). In this regime, a hot convective quasi-spherical shell is formed above the NS magnetosphere, and the plasma entry rate from the shell through the magnetosphere is regulated by the cooling of the hot plasma (due to Compton and radiative energy losses). The hot shell mediates the angular momentum transfer to and from the rotating NS. The equilibrium spin period of NS (at which torques acting on the NS magnetosphere on average vanish) in such a case is mostly determined by the velocity of the stellar wind captured by the NS at the Bondi radius $R_B$ ($v_{rel}$ in Table 1), the orbital binary period $P_b$ , and the neutron star magnetic field (conveniently expressed in terms of the dipole angular momentum $\mu=B_0R_{NS}^3/2$, where $B_0$ is the neutron star surface magnetic field, $R_{NS}$ is the neutron star radius),

\begin{equation}
\label{e:Peq}
P_{eq}\approx 1000[s] \mu_{30}^{12/11} \left(\frac{P_b}{10\,\mathrm{d}}\right)\dot M_{16}^{-4/11}v_8^4
\end{equation}

where $\mu_{30}\equiv\mu/(10^{30}\mathrm{G\,cm}^3)$ and $\dot M_{16}\equiv \dot M/(10^{16}\mathrm{g\,s}^{-1})$, $v_8\equiv v_\mathrm{rel}/(1000\mathrm{\,km\,s}^{-1})$. The stellar wind velocity in 4U 0114$+$65 is about $v_8\sim 0.5-0.7$ (Table 1). For the observed pulse period, Equation (\ref{e:Peq}) implies a high NS magnetic field, of the order of $\mu_{30}\sim 100$, that is, in the magnetar range. This does not seem exotic at
present (see e.g. recent indications on the magnetar nature of a slowly rotating neutron star in the supernova remnant RCW103 \citep{2016arXiv160704107R,2016arXiv160704264D}. 

 In the case of such a strong magnetic field, the Alfv\'{e}n radius of the magnetosphere NS is
 
\begin{equation}
R_A\approx 1.4\times 10^9 [\mathrm{cm}]\mu_{30}^{6/11}\dot M_{16}^{-2/11}\sim 1.7\times 10^{10}[\mathrm{cm}]
\label{e:RA}
\end{equation} 

which is still much smaller than the corotation radius $R_c=(GM_{NS}/\omega^2)^{1/3}\sim 7 \times 10^{10}$~cm. This secures a persistent accretion regime onto the slowly rotating NS even with such a strong magnetic field. The source 4U 0114$+$65 could then be an accreting magnetar \citep{2012MNRAS.425..595R}. This might seem to be at odds with the detection of a cyclotron resonance scattering feature at $E\simeq 22 $keV \citep{bonning}. This detection is uncertain,
however, because other studies \citep{2005astro.ph..8451M, 2006A&A...451..587D,2011MNRAS.413.1083W} have not been able to confirm its existence.

\begin{figure}[]
\centering
\subfigure{\includegraphics[width=\columnwidth]{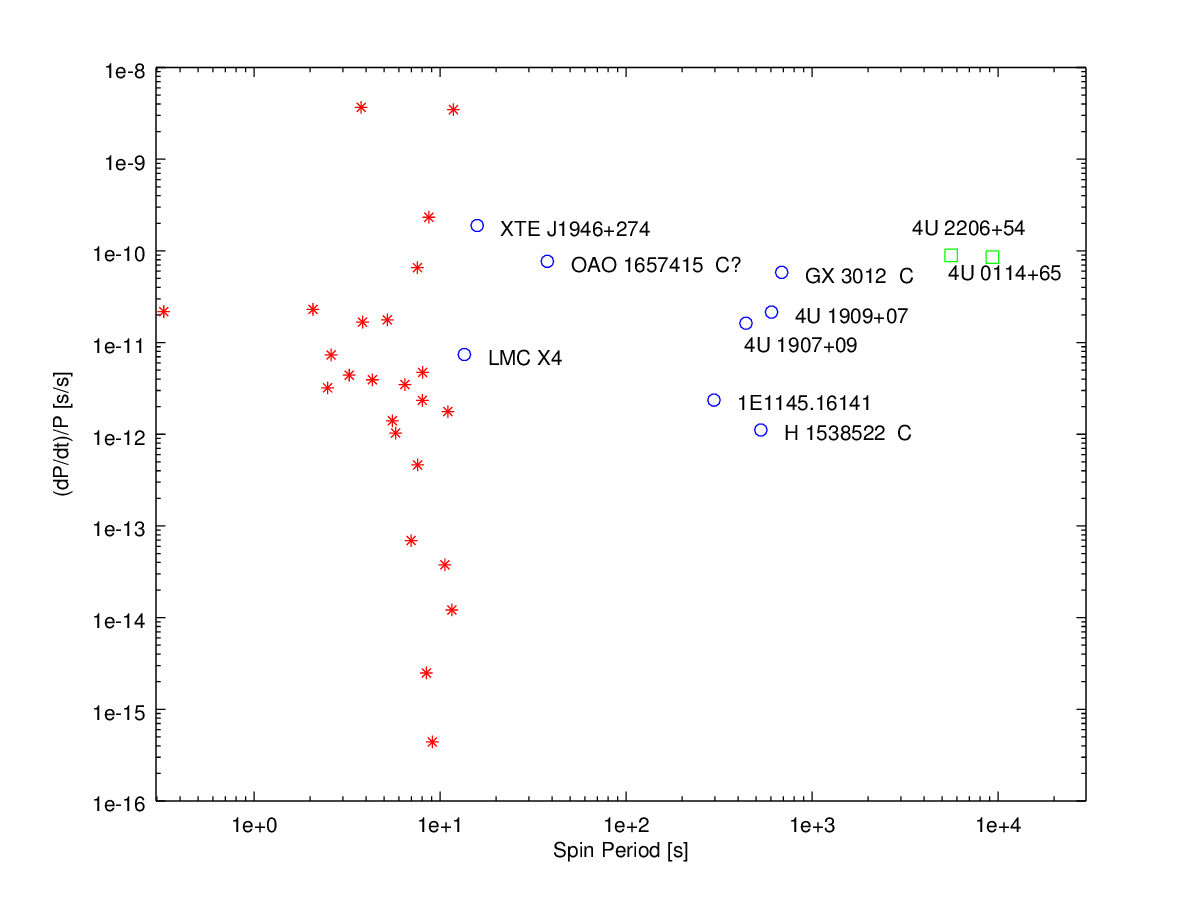}}
\caption{ $ \dot{P}/P$ ( s$^{-1})$ versus $P$ for magnetars \citep[red asterisks;][]{2014ApJS..212....6O} , HMXBs \citep[blue circles;][]{2008A&A...479..533F, 2017arXiv170505205B, 2008A&A...486..293B, 2012arXiv1212.3433S, 2010ApJ...708.1663E,2016MNRAS.458.2745H, 2017A&A...600A..52D, 2012MNRAS.421.2079S}, and accreting magnetar candidates  \citep[green squares; this work;][]{2013MNRAS.432..954W}. The values of  $ \dot{P}/P ( s^{-1})$ are represented in absolute values. Sources labelled with a $C$ present cyclotron lines and are therefore not magnetars.}
\label{spincomp}
\end{figure}

 In Figure \ref{spincomp} we plot the fractional pulse period derivative versus period for magnetars taken from the {\it McGill Magnetar Catalog}  \citep{2014ApJS..212....6O}, along with HMXB pulsars taken from the literature. The sources labelled with a $C$ have cyclotron line detections and are therefore not magnetars. The source 4U 0114$+$65 occupies the same locus as the remaining HMXBs and is close to 4U 2206$+$54, another wind-accreting magnetar candidate. Both sources are at the top of the period derivative distribution. However, non-magnetar sources such as GX301$-$2 also occupy this area. There seems to be no clear segregation in this axis. On the other hand, the spin periods of classical magnetars are evidently typically much shorter than the period of X-ray pulsars, including the two accreting magnetar candidates.

\begin{table*}[htbp]
  \centering
\caption{Compilation of various spectral model parameters used in the literature to fit X-ray spectra of HMXBs,  accreting magnetars (HMXB-AM), and classical magnetars (M). The original paper provides a full description of the models.}
    \begin{tabular}{rrrrrr}
    \toprule
      Source & Model    & $\Gamma$ ($\alpha$)  &$ kT_{\rm col}$/$kT_{\rm bb}$ & Type & Ref.  \\
          &        &    &  keV   \\
\midrule
4U 1909+07 & CompTT      &  & $1.2\pm0.1$ & HMXB & \citet{190907} \\
IGR J18214-1318  & PL+BBODY  & $0.40_{-0.05}^{+0.04}$ & $1.74_{+0.04}^{-0.05}$ & HMXB& \citet{2017ApJ...841...35F}\\
IGR J11215-5952  & PL+BBODY  & $0.56_{-0.11}^{+0.08}$ & $1.61_{+0.06}^{-0.07}$ & HMXB & \citet{2017ApJ...838..133S}\\
LMC X-4  & PL+BREMSS+BBODY  & $0.750\pm 0.005$ & $0.400\pm0.004$ & HMXB&\citet{2017NewA...56...94B} \\
          &        &    &     \\

4U 2206+54 & BMC    & $1.37^{+0.05}_{-0.4}$ & $1.5_{0.4}^{0.7}$& HMXB-AM &\citet{2012MNRAS.425..595R} \\
4U 2206+54 & PL+BB    & $0.94\pm0.03$ & $1.63\pm0.03$&HMXB-AM &\citet{2012MNRAS.425..595R}\\
4U 0114+65 & BMC    & $0.30^{+0.10}_{-0.07}$ &$1.8\pm 0.2$ &HMXB-AM & This work \\
          &        &    &     \\
4U 0142+61 & PL+BB & 3.88(1) &$ 0.41\pm0.10$&M & \citet{2014ApJS..212....6O} \\
SGR 0501+4516 & PL+BB & 3.84(6)& 0.50(2)&M& \citet{2014ApJS..212....6O}\\
SGR 0526-66 & PL+BB & $2.50_{-0.12}^{+0.11}$ & 0.44(2)&M& \citet{2014ApJS..212....6O}\\
1E 1048.1-5937 & PL+BB & 3.14(11) & 0.56(1)&M& \citet{2014ApJS..212....6O}\\
1E 1547.0-5408 & PL+BB & 4.0(2) & 0.43(3)&M& \citet{2014ApJS..212....6O}\\
CXOU J164710.2-455216 & PL+BB & 3.86(22) & 0.59(6)&M& \citet{2014ApJS..212....6O}\\
1RXS J170849.0-400910 & PL+BB & $2.80\pm0.01$ & $0.5_{-0.2}^{+0.4}$&M& \citet{2014ApJS..212....6O}\\
SGR 1806-20 & PL+BB & 1.6(1)& 0.55(7)&M& \citet{2014ApJS..212....6O}\\
1E 1841-045 & PL+BB & 1.9(2)& 0.45(2)&M& \citet{2014ApJS..212....6O}\\
SGR 1900+14 & PL+BB & 1.9(1)& 0.47(2)&M& \citet{2014ApJS..212....6O}\\
SGR 1935+2154 & PL+BB & 1.8(5)& 0.47(2)&M& \citet{2014ApJS..212....6O}\\
1E 2259+586 & PL+BB & 3.75(4)& 0.37(1)&M& \citet{2014ApJS..212....6O}\\
    \bottomrule
    \end{tabular}
  \label{compmod}
\end{table*}

 In an attempt to compare the spectral parameters of different types of sources, we compiled data from the literature in Table \ref{compmod}. Although a meaningful analysis should involve a homogeneous modelling of all sources\footnote{The power-law photon index $\Gamma$ and the \texttt{BMC} spectral index $\alpha$ are not directly comparable. The same is true for the blackbody temperature and the colour temperature of the seed photons. This would require a large-scale reanalysis, which is beyond the scope of this paper.} , some conclusions can be drawn. The seed photon source temperatures of wind-accreting systems (HMXB, including magnetar candidates) is  significantly higher than in classical (non-accreting) magnetars. This might be a consequence of the continuous deposition of energy by the accreted matter in the former.  On the other hand, the spectra of accreting systems are harder on average than in classical magnetars. This might be due to a higher efficiency of the Comptonisation mechanism in accreting systems. However, there seems to be no clear difference between accreting magnetar candidates and the other HMXBs.

{\it The nature of the low-luminosity episodes.} Two episodes of low X-ray emission were observed during our observations. The first was only seen at low energies and coincided with a period of high absorption. This can be interpreted as a stellar wind structure that momentarily  obscures the X-ray source. We can estimate the size of this structure as $l \simeq v_{\rm rel} t_{\rm low}$ , where $v_{\rm rel}^{2}=v^2_{w}+v^2_{orb}$ is the wind velocity relative to the NS and $t_{\rm low}$ is the duration of the low-flux episode. From the analysis of the light curve this turns out to be of the order of $t_{\rm low} \sim 24.6\pm 0.2$ ks ($\lesssim \text{ two}$ pulses). For the relative velocities given in Table 1, this translates into $l\sim[0.30-0.94]\times 10^{12}$ cm $\simeq [0.1-0.4]R_{*}$. This demonstrates that there are large over-dense structures in the supergiant wind, comparable in size with the stellar radius. Figure \ref{emissionlines} shows
thatthe Fe K$\alpha$ line $norm$ is higher during this first episode of low luminosity than during the remaining observation. This suggests that this large over-dense structure is being momentarily illuminated by the NS X-rays, thereby exciting the Fe fluorescence. The presence of such large structures, likely corotating interaction regions (or CIRs),  is commonly invoked to explain the periodic variability observed in the UV spectral lines formed in stellar winds of B-type supergiants \citep{2008ApJ...678..408L} as well as the modulation of X-ray emission from O stars \citep{2001A&A...378L..21O, 2014MNRAS.441.2173M}. 

The interpretation of the second low-luminosity episode is more complex as it affects {\it all energies}. Two mechanisms can cause the pronounced flux decrease: either an increased absorption, or a decrease in mass accretion rate. The first mechanism would require an optically thick structure eclipsing the X-ray source. The duration of the eclipse in the current study is $t_{\rm low} \sim 5.4\pm 0.7$ ks. Therefore, the structure size would be $l\sim[0.6-2.3]\times 10^{11}$ cm $\simeq [0.02-0.09]R_{*}$. If the eclipse were caused by an optically thick structure (unless the obscuring structure is also eclipsing the Fe K$\alpha$ reprocessing region),  we would observe an enhancement in the $EW$ of the Fe K$\alpha$ line as is regularly seen in other HMXBs \citep{2010ApJ...715..947T}. This is very unlikely as this region has an extent larger than $1R_{*}$ (Torrejon et al. 2015).

 The second mechanism involves a cessation of the mass accretion onto the NS. This could  be accomplished in two ways. The first way would be to encounter a region of very low wind density, a wind {\it bubble} \citep{2005A&A...432..999V}. 
This also would contradict a nearly constant or even  slightly increasing absorbing column $N_{\rm H}$ as deduced from the observations. Therefore, a wind bubble seems unlikely. The second mechanism could be centrifugal inhibition of accretion. This {\it propeller} mechanism originally suggested by \citep{1975A&A....39..185I} is now frequently  invoked to explain observed low-luminosity episodes in some X-ray pulsars (see e.g. the recent paper of \citep{2017ApJ...834..209L} and references therein). It could also be responsible for many of the {\it off states} (periods in which the accretion is totally halted) seen in some HMXBs \citep[cf.][]{2008A&A...492..511K}. However, the NS in 4U 0114$+$65 has a spin period of 2.6 h, which, as we showed, renders the propeller mechanism ineffective ($R_A<<R_c$) even for magnetar fields.  For a propeller to start operating, the Alfv\`{e}n radius $R_A$ should become comparable to the corotation radius $R_c$. This would require a drop of 1000 times in the X-ray luminosity.
An alternative explanation of the observed low-luminosity episode could involve a substantial departure from the spherical accretion symmetry. The Bondi radius is an idealisation and could change  by a factor 2 or 3 in different directions, as the stellar wind  velocity changes on the characteristic scale of $10^{11}$ cm, which is comparable to $R_B$ (see Table 1). The stellar wind could then be momentarily captured in a non-spherical way. Since the magnetospheric radius and the Bondi radius differ only by a factor of a few, the captured matter can reach the magnetosphere non-spherically, thereby enabling a switching between the magnetic poles of the NS. This pole switching can also be responsible for the pulse shape change displayed by the source between T1 and T2. 

\section{ Summary and conclusions}

We have performed a time-resolved investigation of one of the slowest pulsars ever found using a 49 ks uninterrupted observation with the {\it XMM-Newton}.  The following conclusions can be drawn from our analysis:

\begin{enumerate}

\item We successfully fit the pulse-phase resolved spectra with a bulk motion Comptonisation model (\texttt{bmc}) for the first time. This model implies a very  small ($r\sim 3$ km) and hot ($kT\sim 2-3$ keV) emitting region and therefore points to a hot spot on the surface of a rotating NS as the most likely explanation for the X-ray pulse. 

\item The long NS spin period ($P_{spin}\sim 9.4$ ks) and the secular spin-up $\dot{\nu}=(8.5  \pm 0.8) \times 10^{-15}$ Hz s$^{-1}$ can be explained within the theory of quasi-spherical settling accretion onto the NS provided that the magnetic field is in the magnetar range $\mu_{30}\sim 30-100$. 

\item Owing to the extremely slow NS rotation period, even with such a strong magnetic field, the magnetospheric radius ($R_{A}$) is still much smaller than the corotation radius $R_{c}$. This is consistent with the persistent nature of this source.

\item The stellar wind of the supergiant star presents a covering fraction ($\sim 0.3$) much lower than  typically found for other systems with supergiant donors ($\sim 0.8-0.9$). This indicates a much lower degree of wind clumping. This is also supported by the much weaker Fe K$\alpha$ line as compared to other HMXBs.
 We suggest that the proximity of 4U 0114$+$65 to the bistability jump, where the mass loss and the wind acceleration change dramatically, could have a strong impact on wind clump formation.

\item The light curve presents dips that are clearly seen within the HF states. The short dips tend to show a colour ratio increase and would be caused by wind clump absorption. The long dips, in turn, tend to show a constant CR and would be produced by a general decrease in the mass accretion rate associated with the pass of an interclump zone by the NS. Its characteristic sizes would be $\simeq 1.6 \times 10^{-3} R_{*}$ ($4 \pm 3\times 10^{9}$ cm)  and $\simeq 1.2 \times 10^{-2} R_{*}$ ($2.9 \pm 2.2 \times 10^{10}$ cm) for the long and short dips, respectively. The interclump to clump size ratio observed is $\sim 7,$ which
is close to the expected value ($\sim 10$) predicted by the theoretical models of stellar winds in massive stars.

\item We detected two episodes of low luminosity. The first affects only the low energies and can be attributed to a large ($l\sim [0.3-0.5]R_{*}$) over-dense structure in the B1 supergiant donor wind, compatible with a corotating interaction region (CIR). The second episode affects all energies. An eclipse due to an optically thick structure or accretion failure caused by a rarefied cavity in the wind seems to be unlikely. Considering the small difference (within a factor of one) between the Bondi and magnetospheric radii, spherically asymmetric capture of matter close to the Bondi radius might cause temporal cessation of accretion onto one NS magnetic pole. Such a mass accretion redistribution between the NS poles could also produce the observed change in the NS pulse shape. 

\end{enumerate}

\begin{acknowledgements}
This work has been funded by the research grants ESP2014-53672-C3-3P and 50 OR 1508 (LO). We acknowledge the detailed comments of the anonymous referee that improved the content and presentation of the paper. The authors acknowledge the hospitality of the ISSI in Bern under the team lead by S. Martinez.
\end{acknowledgements}

%BIBLIOGRAFIA
%\bibliographystyle{aa}
\bibliographystyle{plainnat}
\bibliography{bib}

\newpage
\onecolumn

\begin{appendix}

\section{Phase-resolved spectra for the \texttt{bmc} model. }
\begin{table*}[!h]
\centering
\caption{Phase-resolved parameters for the spectral model.}
\begin{tabular}{rrrrrrrr}
\toprule
 & $ \chi ^2 $ & $C$ & $N_{H}^{1}$ & $kT$ & $\alpha$ & $norm$ & $N_{H}^{2}$ \\
\textbf{} & & & $\times10^{22} cm^{-2}$ & keV & \texttt{bmc} & $\times 10^3$ &$\times10^{22} cm^{-2}$ \\
\midrule
AS & 1,39 & $0.38^{+0.17}_{-0.15}$ & $0.56^{+0.12}_{-0.14}$ & $1.8^{+0.02}_{-0.02}$ & $0.3^{+0.10}_{-0.07}$ & $5.2^{+1.0}_{-0.8}$ & $1.5^{+0.3}_{-0.2}$ \\\\
HF1 & 1.01 & $0.1^{+0.01}_{-0.10}$ & $0.47^{+0.05}_{-0.03}$ & $1.83^{+0.05}_{-0.01}$ & $0.27^{+0.01}_{-0.01}$ & $7.5^{+4}_{-3}$ & $2.52^{+0.14}_{-0.13}$ \\\\
LF1 & 1.40 & $0.1^{+0.02}_{-0.01}$ & $0.56^{+0.17}_{-0.12}$ & $1.65^{+0.05}_{-0.05}$ & $0.07^{+0.01}_{-0.01}$ & $3.34^{+0.07}_{-0.07}$ & $5.1^{+0.3}_{-0.3}$ \\\\
HF2 & 1.10 & $0.05^{+0.01}_{-0.01}$ & $0.49^{+0.06}_{-0.04}$ & $1.88^{+0.06}_{-0.04}$ & $0.2^{+0.5}_{-0.1}$ & $8.82^{+0.05}_{-0.05}$ & $6.05^{+0.07}_{-0.09}$ \\\\
LF2 & 1.14 & $0.10^{+0.03}_{-0.01}$ & $0.68^{+0.12}_{-0.14}$ & $1.68^{+0.10}_{-0.09}$ & $1.6^{+0.7}_{-0.4}$ & $0.7^{+0.01}_{-0.01}$ & $2.26^{+0.11}_{-0.11}$ \\\\
HF3 & 1.09 & $0.1^{+0.01}_{-0.01}$ & $0.48^{+0.03}_{-0.02}$ & $1.72^{+0.02}_{-0.02}$ & $0.31^{+0.01}_{-0.01}$ & $7.5^{+0.04}_{-0.04}$ & $2.06^{+0.03}_{-0.03}$ \\\\
LF3 & 1.10 & $0.30^{+0.04}_{-0.05}$ & $0.66^{+0.14}_{-0.04}$ & $1.30^{+0.03}_{-0.04}$ & $0.1^{+0.01}_{-0.01}$ & $2.02^{+0.04}_{-0.04}$ & $1.75^{+0.12}_{-0.13}$ \\\\
HF4 & 1.55 & $0.4^{+0.4}_{-0.4}$ & $0.66^{+0.14}_{-0.04}$ & $1.59^{+0.01}_{-0.01}$ & $0.31^{+0.01}_{-0.01}$ & $8.13^{+0.11}_{-0.04}$ & $0.67^{+0.02}_{-0.02}$ \\\\
LF4 & 1.04 & $0.17^{+0.11}_{-0.01}$ & $0.42^{+0.11}_{-0.01}$ & $1.24^{+0.03}_{-0.03}$ & $0.3^{+0.01}_{-0.02}$ & $1.19^{+0.01}_{-0.01}$ & $0.85^{+0.06}_{-0.02}$ \\\\
HF5 & 1.64 & $0.14^{+0.7}_{-0.04}$ & $0.79^{+0.01}_{-0.06}$ & $1.48^{+0.01}_{-0.01}$ & $0.31^{+0.01}_{-0.01}$ & $3.35^{+0.01}_{-0.01}$ & $0.78^{+0.01}_{-0.01}$ \\\\
LF5a & 1.10 & $0.19^{+0.08}_{-0.04}$ & $0.64^{+0.11}_{-0.04}$ & $1.16^{+0.03}_{-0.03}$ & $0.35^{+0.03}_{-0.02}$ & $0.94^{+0.01}_{-0.01}$ & $1.19^{+0.07}_{-0.05}$ \\\\
LF5b & 1.14 & $0.21^{+0.11}_{-0.11}$ & $0.80^{+0.01}_{-0.13}$ & $1.21^{+0.05}_{-0.04}$ & $0.42^{+0.04}_{-0.04}$ & $0.39^{+0.01}_{-0.01}$ & $1.18^{+0.07}_{-0.07}$ \\\\
HF6 & 1.25 & $0.22^{+0.05}_{-0.12}$ & $0.68^{+0.01}_{-0.04}$ & $1.42^{+0.02}_{-0.02}$ & $0.30^{+0.01}_{-0.01}$ & $3.67^{+0.03}_{-0.03}$ & $0.99^{+0.04}_{-0.03}$ \\\\
\bottomrule
\end{tabular}
\label{bmctable}
\end{table*}

\begin{figure*}[!htbp]
\centering
\caption{Spectra for the BMC model. The panels are in chronological order.}
\begin{tabular}{rrrrr}
\subfigure{\includegraphics[width=40mm]{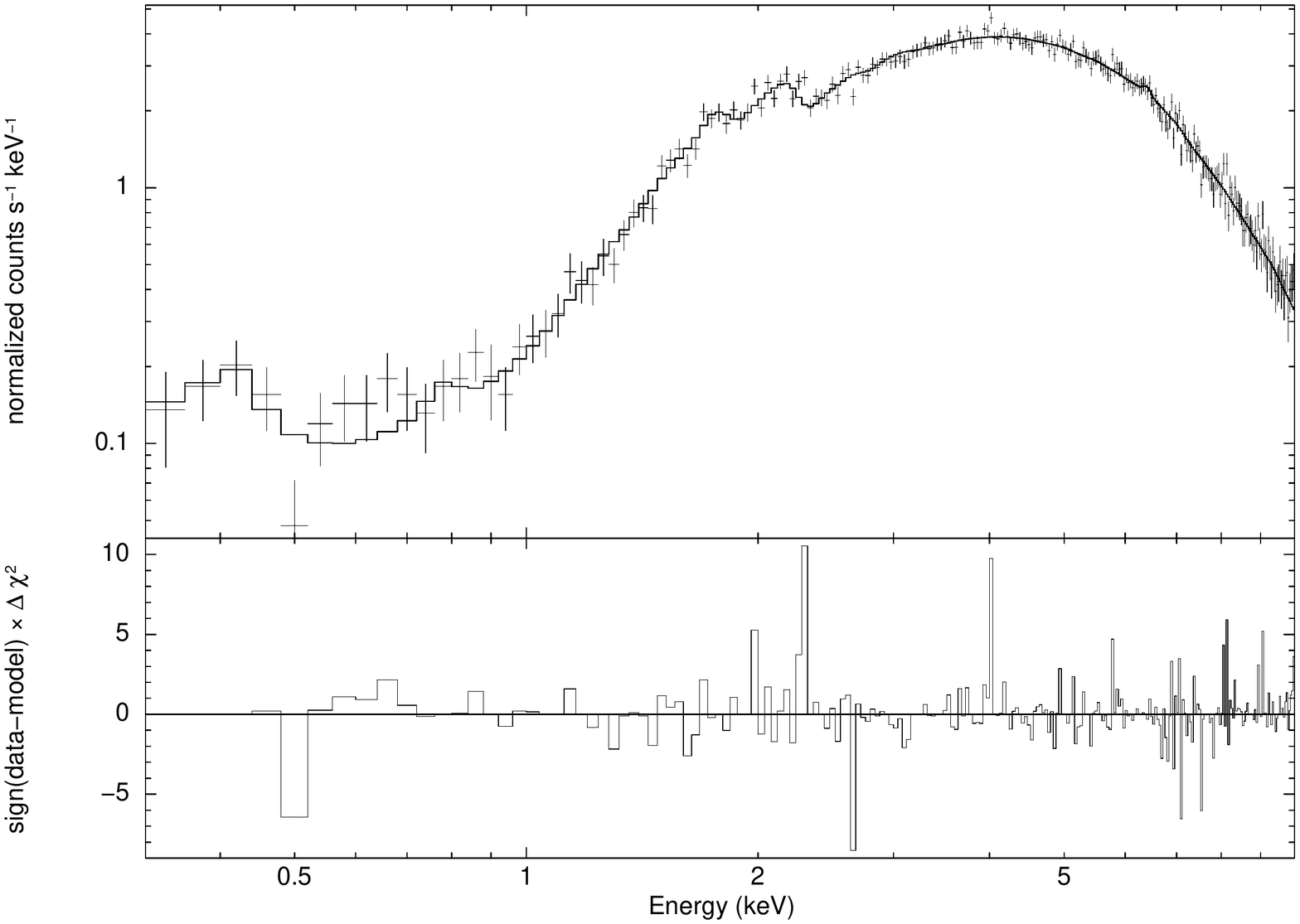}}&
\subfigure{\includegraphics[width=40mm]{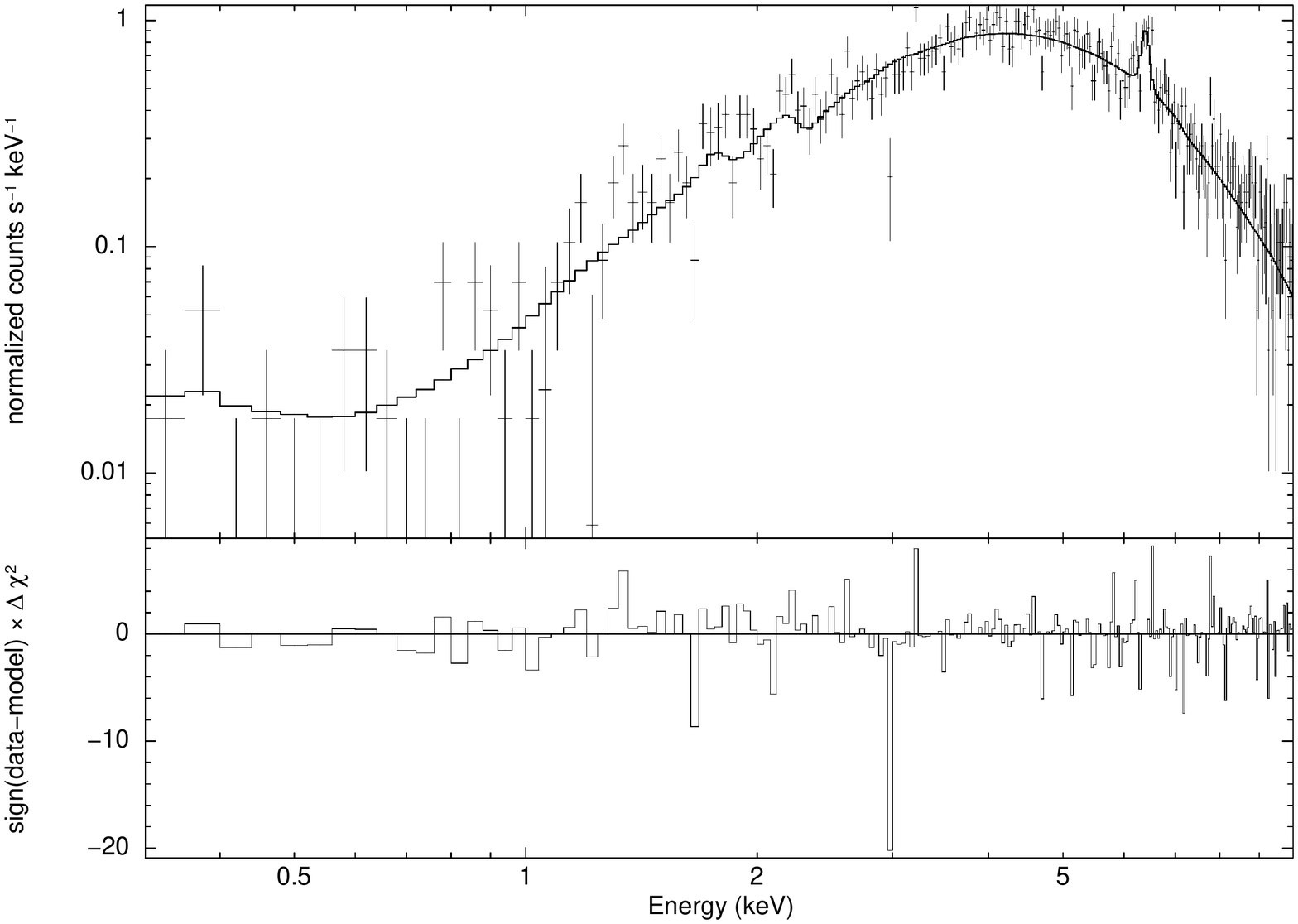}}&
\subfigure{\includegraphics[width=40mm]{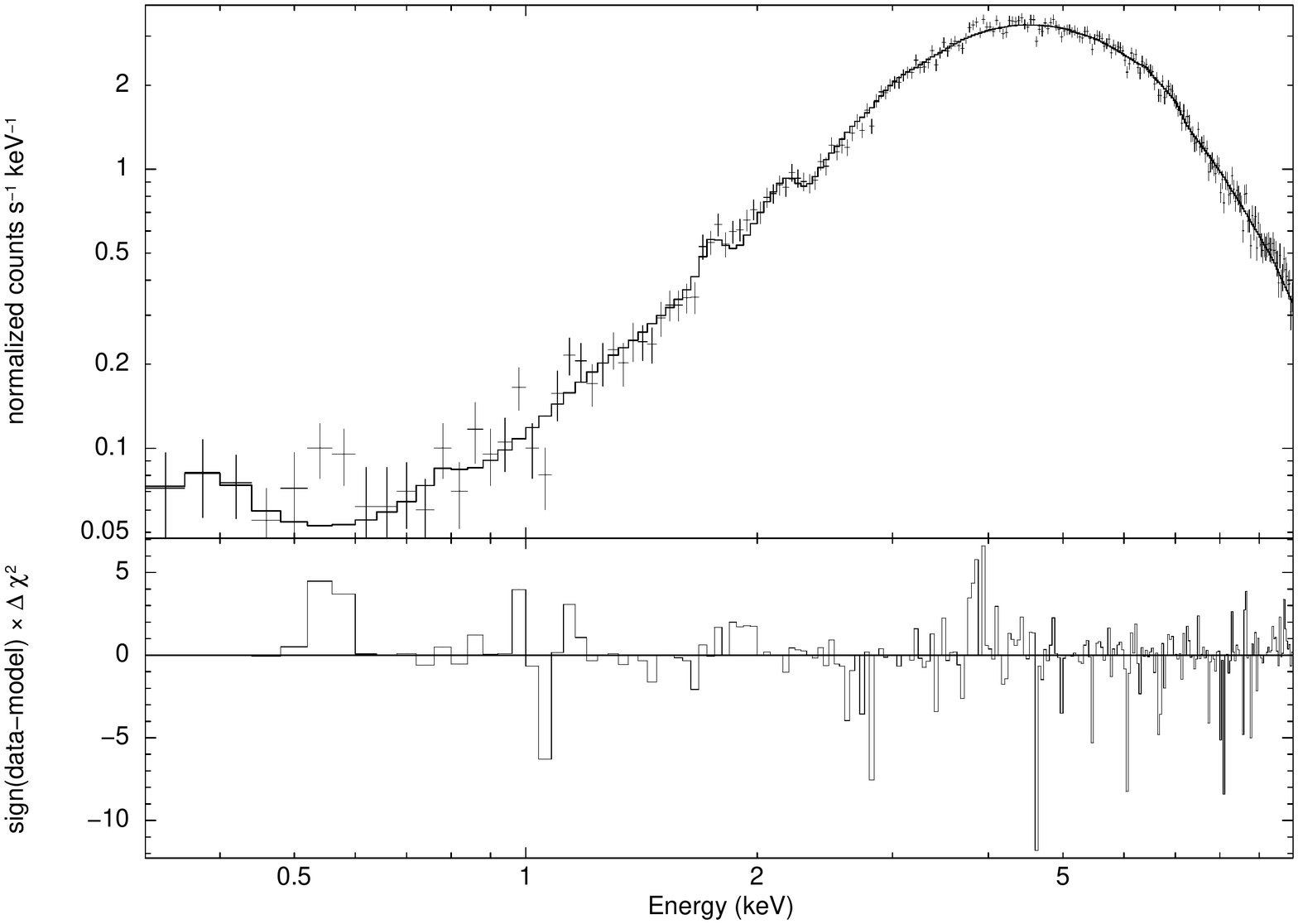}}&
\subfigure{\includegraphics[width=40mm]{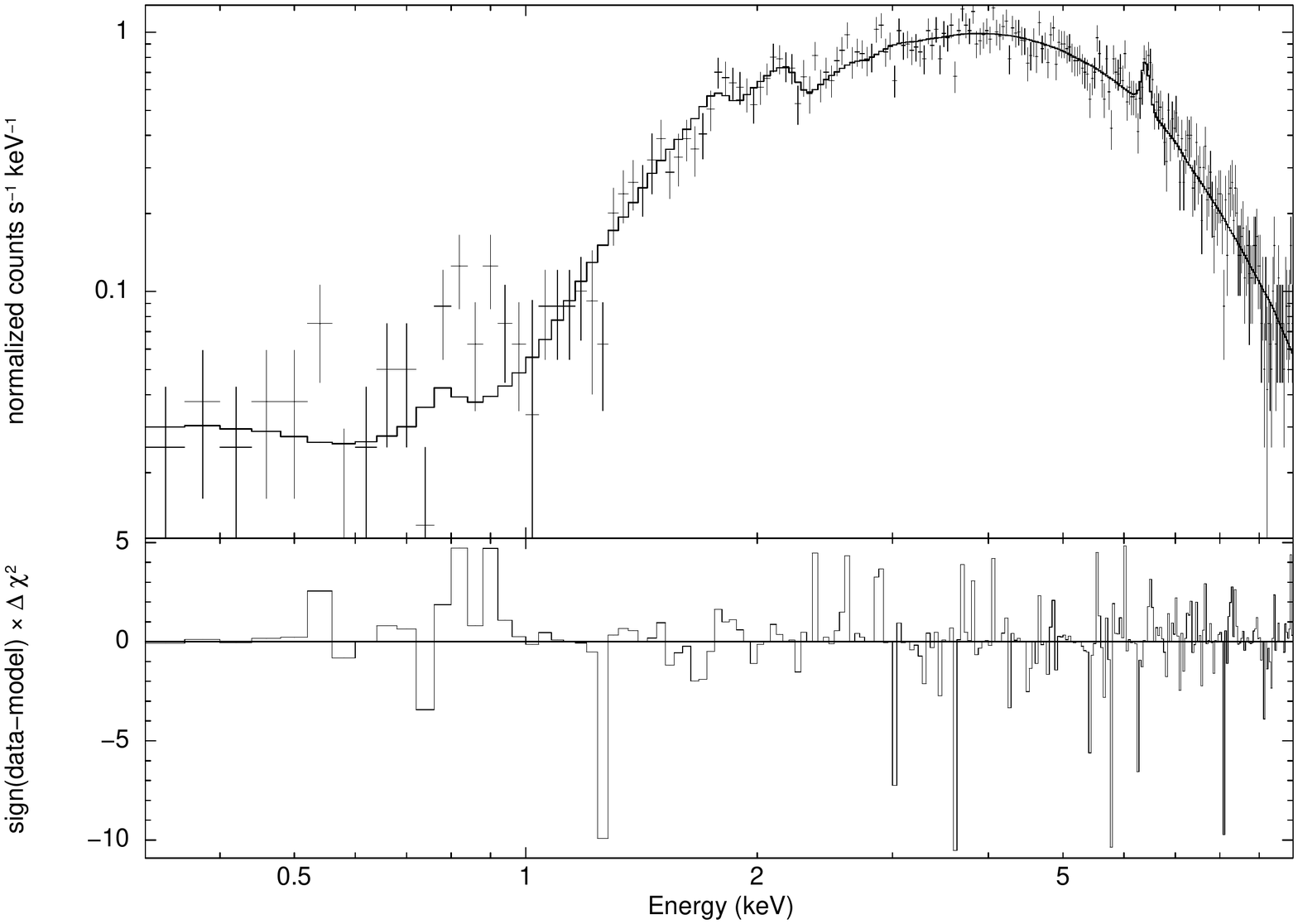}}\\
\subfigure{\includegraphics[width=40mm]{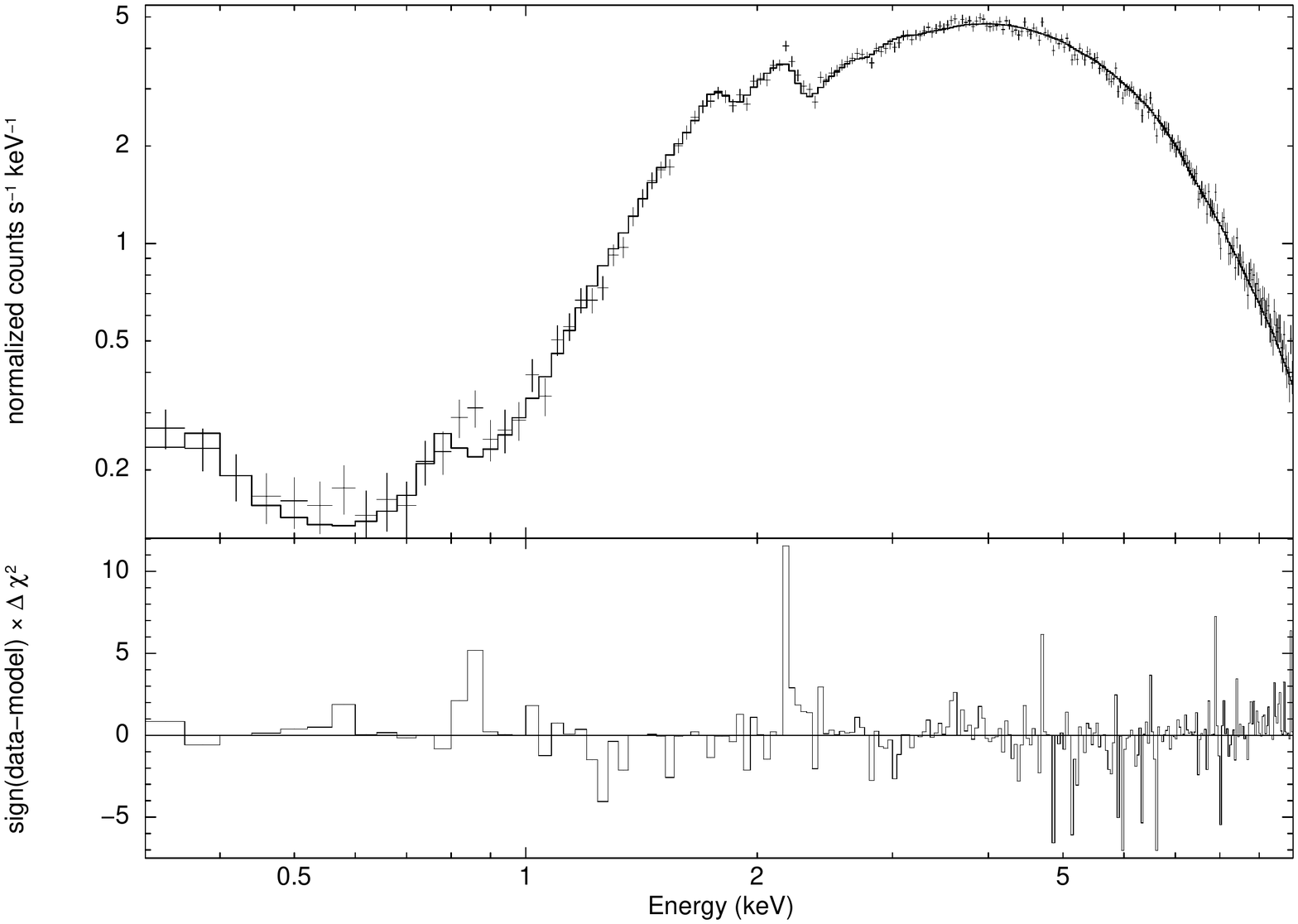}}&
\subfigure{\includegraphics[width=40mm]{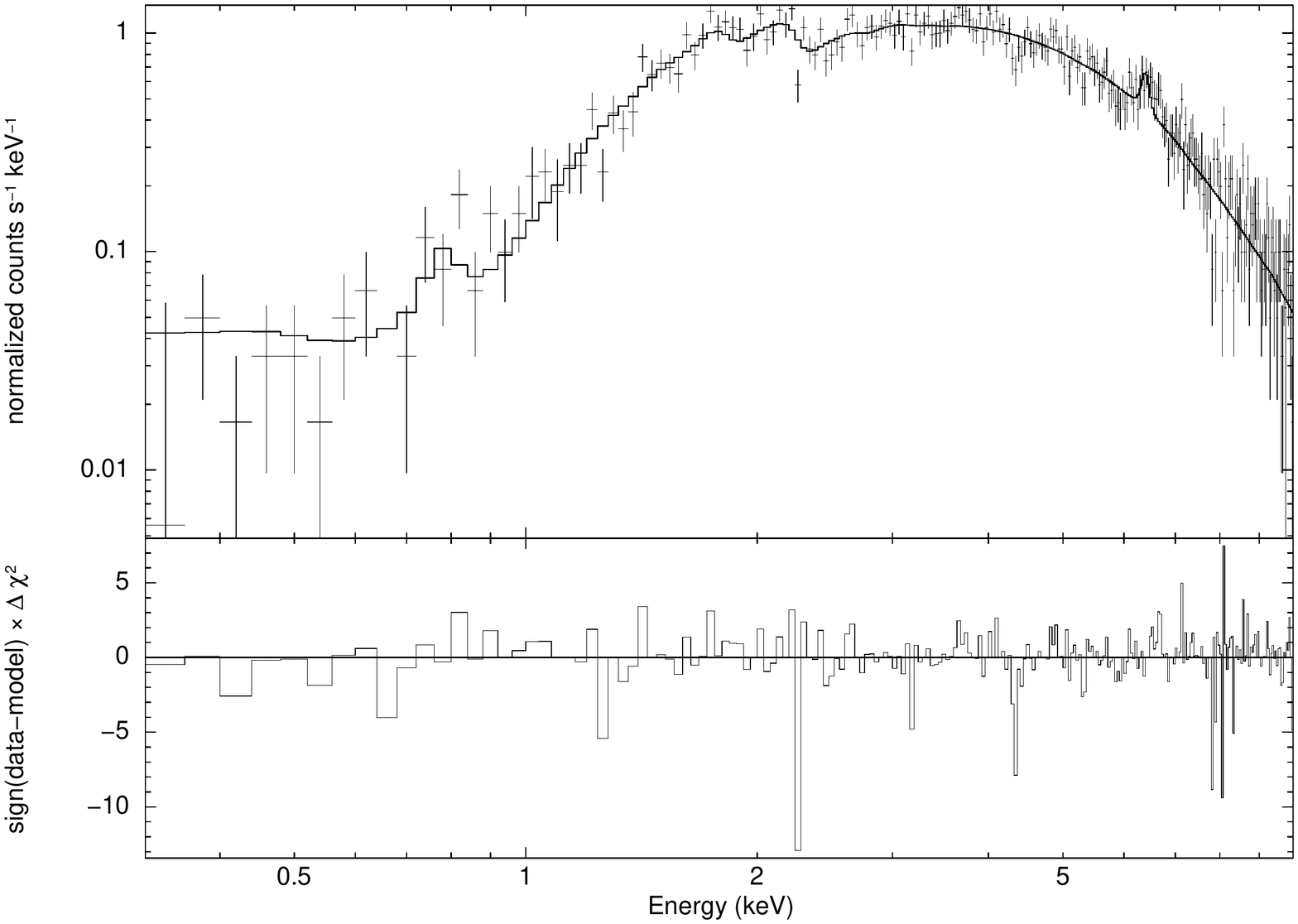}}&
\subfigure{\includegraphics[width=40mm]{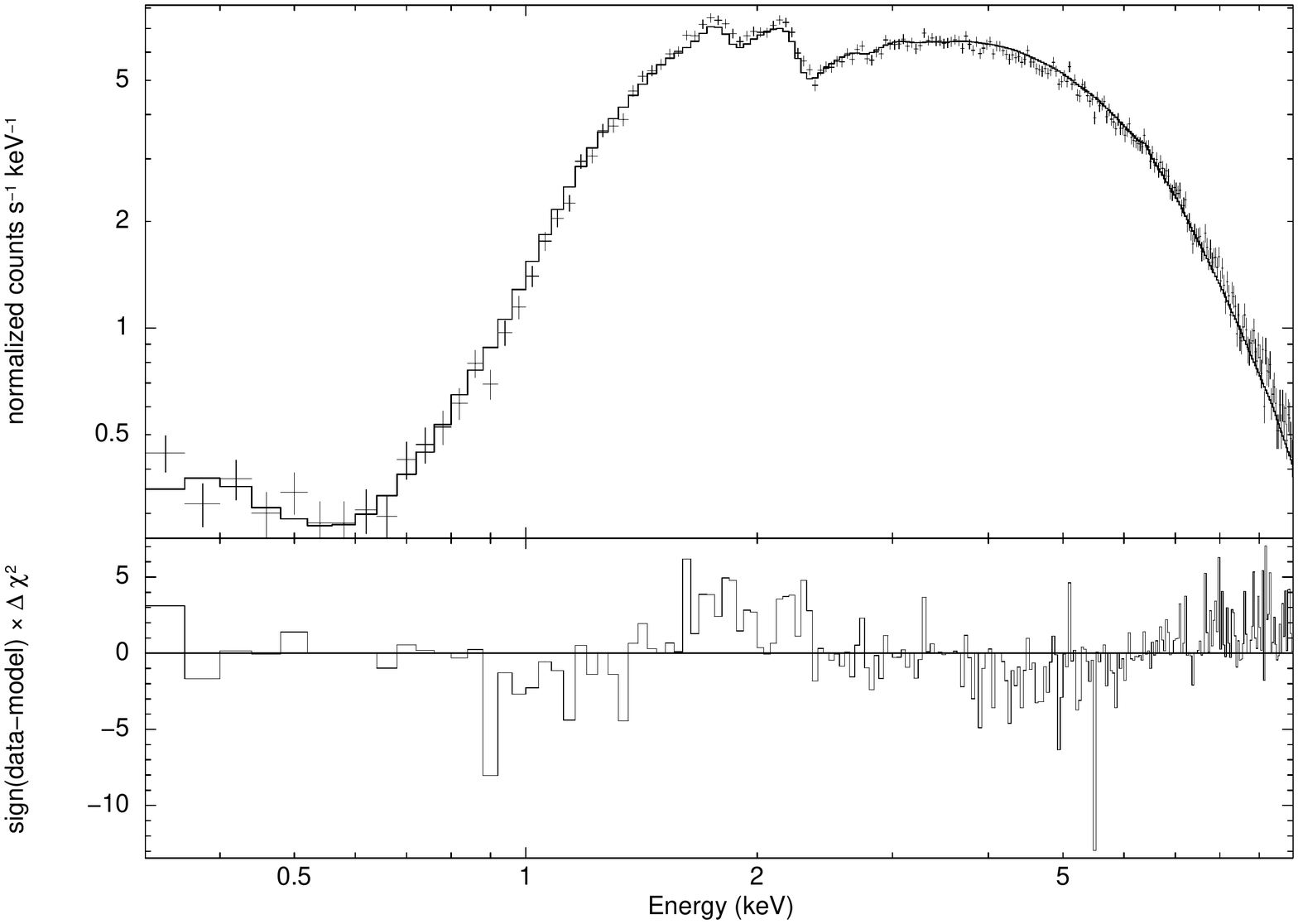}}&
\subfigure{\includegraphics[width=40mm]{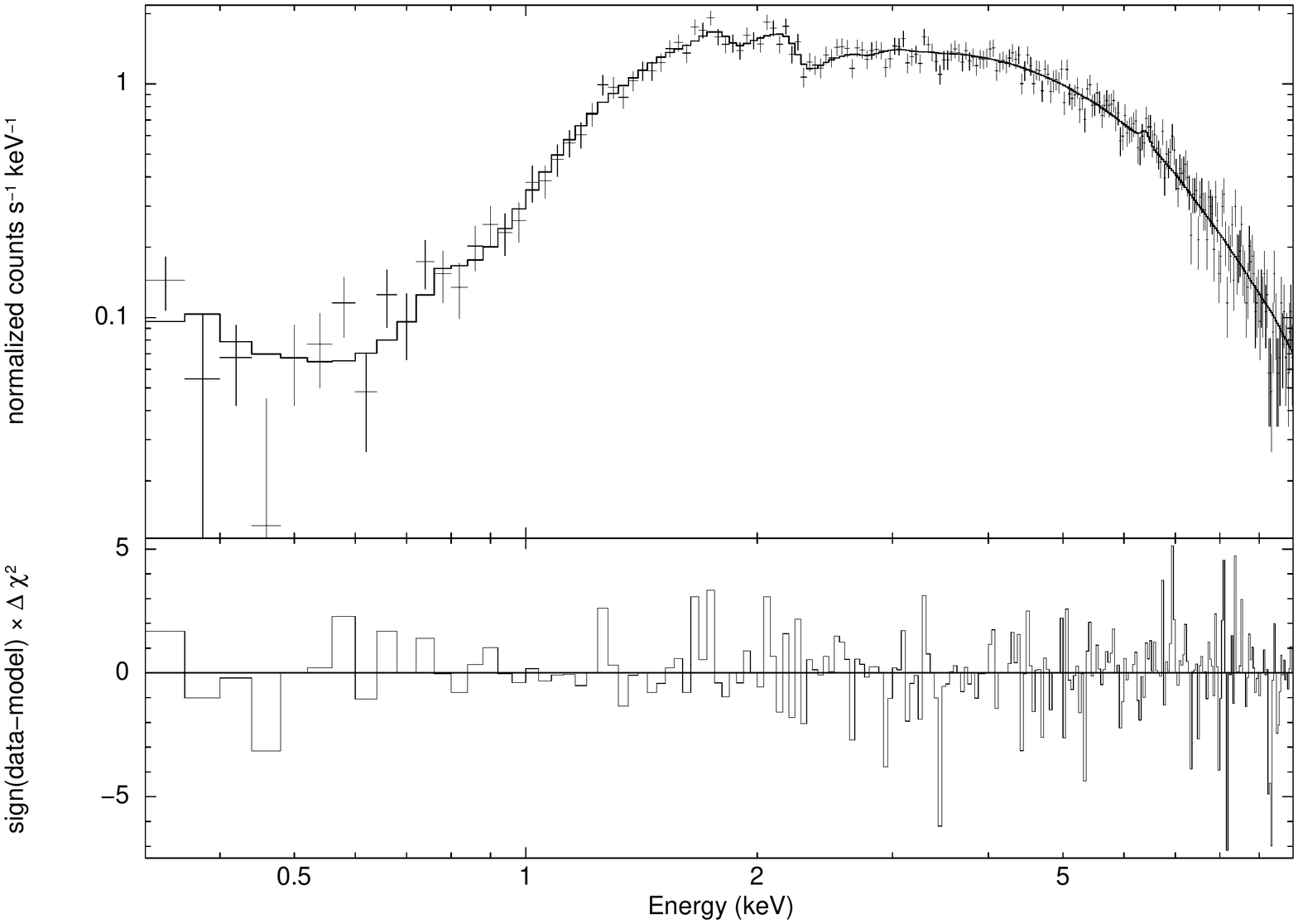}}\\
\subfigure{\includegraphics[width=40mm]{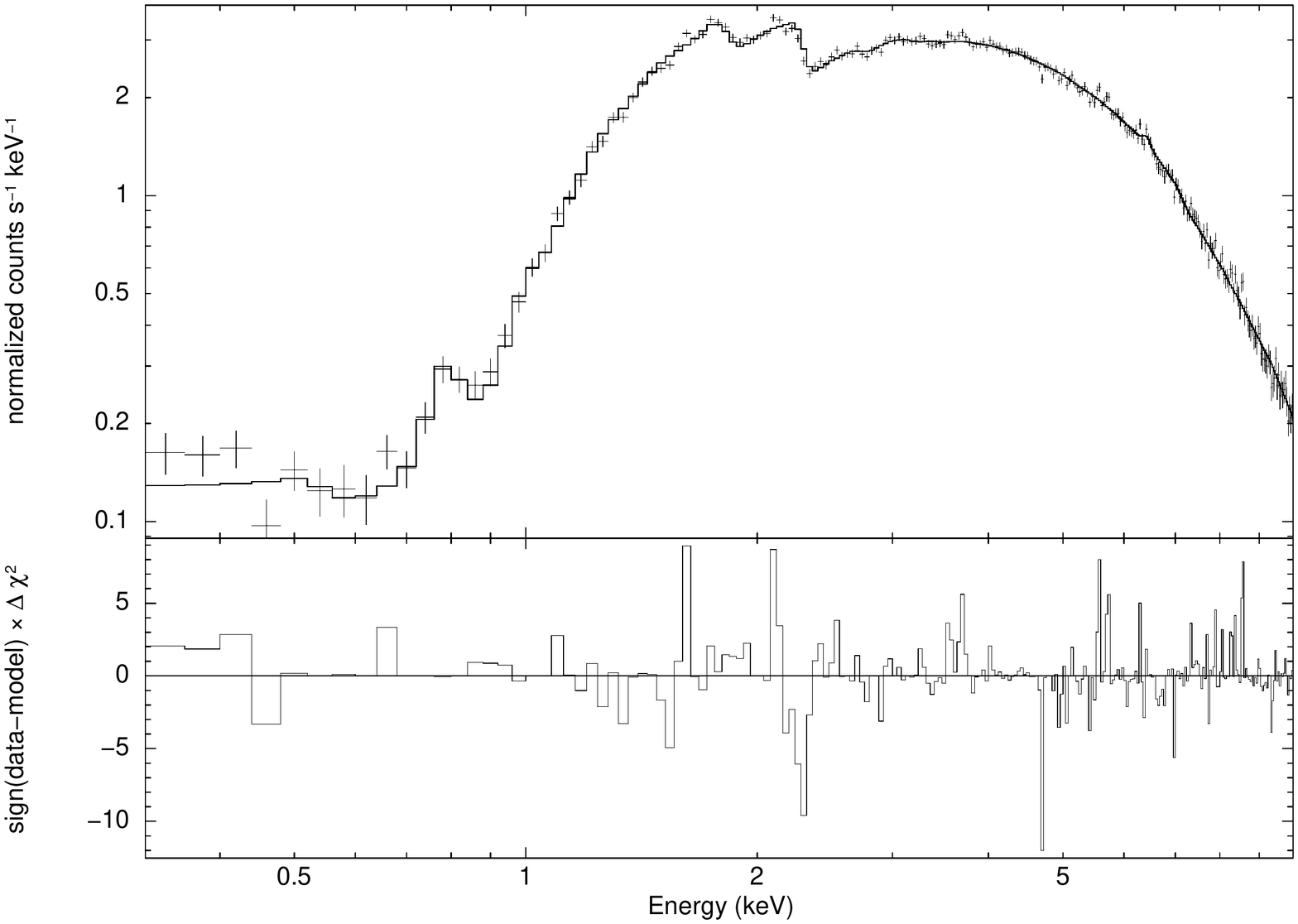}}&
\subfigure{\includegraphics[width=40mm]{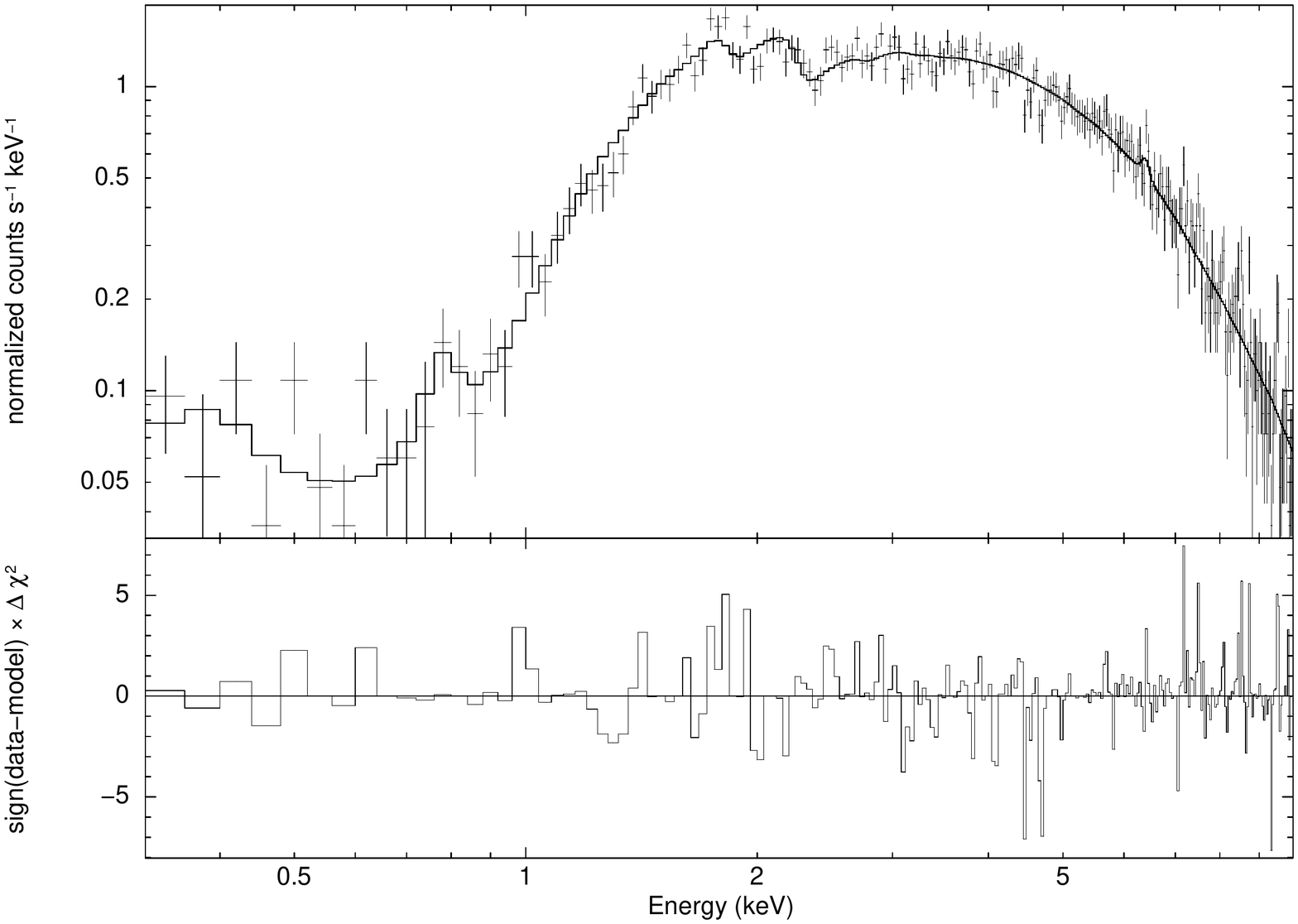}}&
\subfigure{\includegraphics[width=40mm]{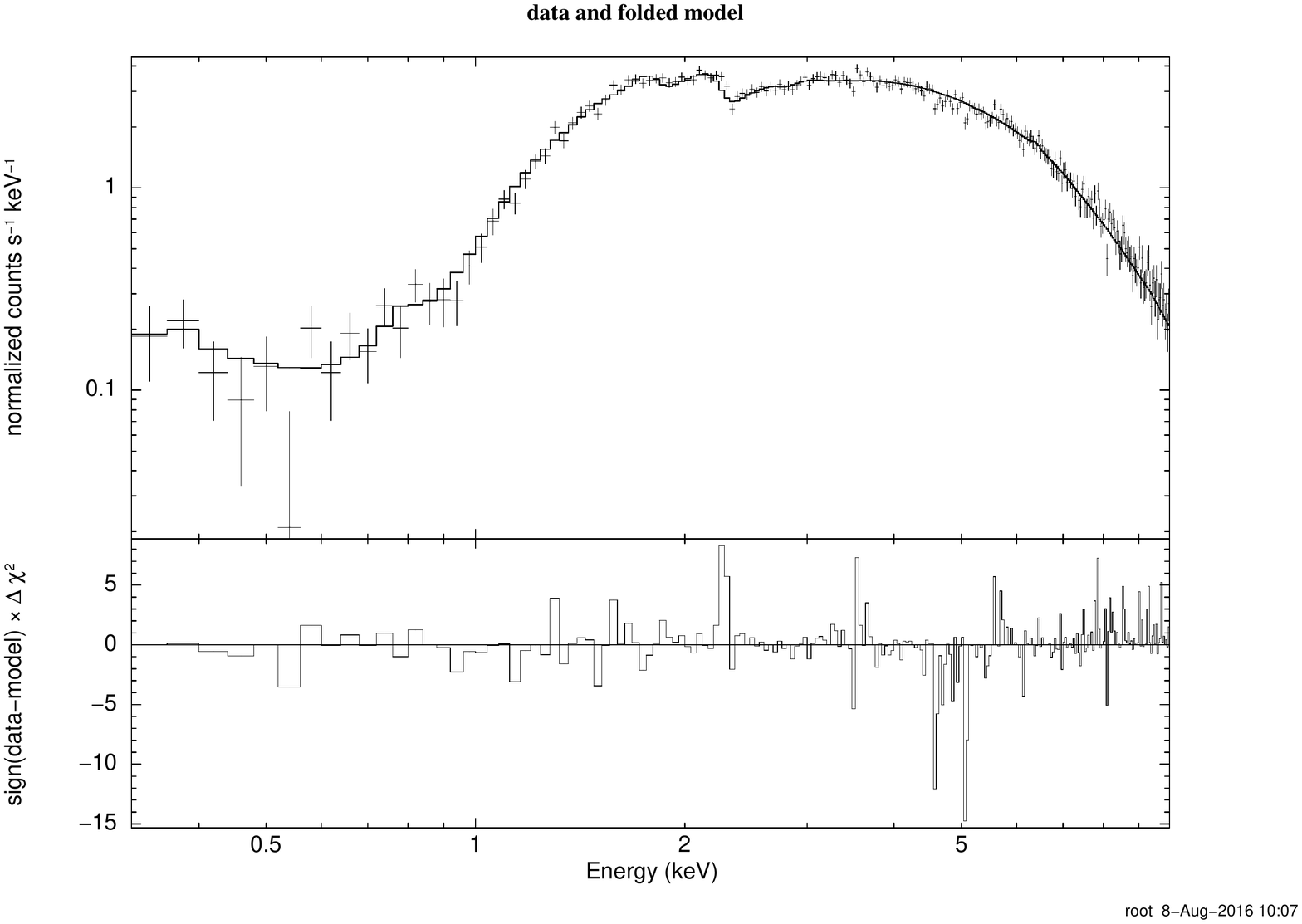}}\\
\label{appendix:A}
\end{tabular}
\end{figure*}

\newpage
\section{Phase-resolved analysis of the emission line parameters. }
\begin{table*}[!h]
\centering
\caption{Phase-resolved analysis of the emission lines. In some instances (labelled with an asterisk), only upper limits can be given, while in others the detection was null. }
\begin{adjustbox}{max width=\textwidth}
\begin{tabular}{rrrrr}
\toprule
Energy & 6.4 keV & & 0.78 keV & \\
\midrule
& $norm$ & $EW$ & $norm$ & $EW$ \\
&$\times 10^{-5} $&$\times 10^{-3}$  keV & $\times 10^{-5} $& $\times 10^{-2} $ keV  \\
\midrule

S & $2.0^{+0.7}_{-0.7}$ & $3.9^{+1.3}_{-1.3}$ & $9.8^{+7}_{-5}$ & $5.9^{+4}_{-3}$ \\\\

HF1 & $*3^{+4}_{-3}$ & $1.3^{+1.4}_{-1.3}$ & $*20^{+40}_{-20}$ & $11^{+17}_{-11}$ \\\\

LF1 & $9.1^{+2.4}_{-2.4}$ & $14^{+4}_{-4}$ &... &... \\\\

HF2 & $*1.3^{+2.4}_{-1.3}$ & $0.3^{+0.5}_{-0.3}$ &...&... \\\\

LF2 & $5.5^{+1.9}_{-1.9}$ & $9^{+3}_{-3}$ & $*15^{+50}_{-15}$ & $*30^{+80}_{-30}$ \\\

HF3 &...&...& $50^{+30}_{-30}$ & $18^{+11}_{-11}$ \\\\

LF3 & $4.4^{+2.0}_{-2.0}$ & $24^{+11}_{-11}$ & $*20^{+30}_{-20}$ & $*25^{+30}_{-24}$ \\\\

HF4 & $*3^{+3}_{-3}$ & $*3^{+4}_{-3}$ &...&... \\\\

LF4 & $*1.3^{+1.5}_{-1.3}$ & $*4^{+4}_{-4}$ & $*6^{+8}_{-6}$ & $*3^{+3}_{-3}$ \\\\

HF5 & $2.0^{+1.1}_{-1.1}$ & $1.8^{+1.0}_{-1.0}$ & $16^{+8}_{-8}$ & $0.6^{+0.3}_{-0.3}$ \\\

LF5a &...&...& $28^{+24}_{-24}$ & $10^{+9}_{-9}$ \\\\

LF5b &$*0.3^{+0.8}_{-0.3}$ & $*2.4^{+7}_{-2.4}$ & $*6^{+15}_{-6}$ & $*4^{+11}_{-4}$ \\\\

H6 & $*1.5^{+2.0}_{-1.5}$ & $*2^{+3}_{-2}$ & $*21^{+24}_{-21}$ & $*3^{+4}_{-3}$ \\\\
\bottomrule
\end{tabular}
\end{adjustbox}
\label{appendix:C}
\end{table*}

\end{appendix}

\end{document}